\documentclass[superscriptaddress,aps,showpacs,nofootinbib,twocolumn]{revtex4}
\usepackage{graphicx}
\usepackage{amsmath,amssymb}

\def\vev#1{\langle {#1}\rangle}

\def\frac#1#2{{\textstyle{{#1}\over {#2}}}}

\def\lsim{\mathrel{\rlap{\lower4pt\hbox{\hskip1pt$\sim$}}
    \raise1pt\hbox{$<$}}}
\def\gsim{\mathrel{\rlap{\lower4pt\hbox{\hskip1pt$\sim$}}
    \raise1pt\hbox{$>$}}}
\def\sqr#1#2{{\vcenter{\vbox{\hrule height.#2pt
         \hbox{\vrule width.#2pt height#1pt \kern#1pt
         \vrule width.#2pt}
         \hrle height.#2pt}}}}

\def\beq{\begin{equation}}
\def\eeq{\end{equation}}
\def\beqa{\begin{eqnarray}} 
\def\eeqa{\end{eqnarray}}

\def\laq{\raise 0.4 ex \hbox{$<$}\kern -0.8 em\lower 0.62 ex\hbox{$\sim$}}
\def\gaq{\raise 0.4 ex \hbox{$>$}\kern -0.7 em\lower 0.62 ex\hbox{$\sim$}}

\begin{document}

\title{Resonant particle production in branonium}

\author{J. G. Rosa}

\email{j.rosa1@physics.ox.ac.uk}

\author{John March-Russell}

\email {jmr@thphys.ox.ac.uk}

\affiliation{The Rudolph Peierls Centre for Theoretical Physics,\\ Department of Physics, University of Oxford,\\ 1 Keble Road, Oxford OX1 3NP, UK}


\date{\today}

\begin{abstract}
We study the mechanism of particle production in the world-volume of a probe $\overline{D6}$-brane (or $D6$ with SUSY breaking) moving in the background created by a fixed stack of $D6$-branes. We show that this may occur in a regime of parametric resonance when the probe's motion is non-relativistic and it moves at large distances from the source branes in low eccentricity orbits. This leads to an exponential growth of the particle number in the probe's world-volume and constitutes an effective mechanism for producing very massive particles. We also analyze the evolution of this system in an expanding universe and how this affects the development of the parametric resonance. We discuss the effects of transverse space compactification on the probe's motion, showing that it leads to the creation of angular momentum in a similar way to the Affleck-Dine mechanism for baryogenesis. Finally, we describe possible final states of the system and their potential relevance to cosmology.
\end{abstract}


\maketitle


\section{Introduction}

One of the most important aspects of string theory with potential applications to beyond-the-standard-model physics and cosmology is the existence of additional extended objects.  A particularly interesting class of these, $Dp$-branes, occur as $(p+1)$-dimensional surfaces arising as supersymmetric solutions of the low-energy supergravity field equations but have also a high-energy interpretation as surfaces confining the endpoints of open strings \cite{Polchinski}.   By now many different configurations of $Dp$-branes have been
studied, including parallel and intersecting sets of branes \cite{Berkooz,SMIntersectingBranes}, the combination of branes with different dimensionalities \cite{Douglas} and also several possible embeddings in the compactified extra-dimensions \cite{embedding}. These models have an enormous potential in the construction of gauge theories with matter and several new extensions of the Standard Model have been obtained in this
way \cite{SMIntersectingBranes}.

Branes have also been intensively studied in their application to the problems of early universe cosmology, such as inflation and reheating. In the context of effective field theory treatments of brane-world model building, an immense amount of work has been performed on the astrophysical and cosmological implications of these theories for both flat and warped extra dimensions \cite{earlybranecosmo}, while many studies in string theory have also been undertaken \cite{Quevedo,Martineau,braneinflation}. One important property of $Dp$-branes is the fact that they interact at large distances via the exchange of closed string modes, which allows one to study them as massive point-like charged objects moving along the dimensions transverse to their world-volumes.  An interesting application of this was proposed in \cite{Burgess}, where it was shown that branes with opposite charges may form bound states similar to electron-positron states, despite the different origin of the central potentials involved. This analogy has motivated the name of \textit{branonium} to designate this type of systems. Several properties of these bound states have been analyzed in \cite{Burgess}, including the classical orbits, the associated quantum dynamics and possible compactification schemes. This analysis assumed a probe (anti)brane moving in the background spacetime created by a fixed central stack of branes, thus neglecting the gravitational backreaction of the probe. In the case of $D6$-branes, the probe's trajectory follows closed elliptical orbits similar to those found in planetary systems. It was also suggested that the probe's motion would lead to radiation of particles into both the bulk space and the probe's world-volume, making otherwise stable orbits decay. The stability of this system was further studied in \cite{Lukas}, where it was shown that orbital decay necessarily occurs when geometric moduli of the background spacetime are allowed to vary.

In this work, we explore some properties of branonium systems, namely the mechanism leading to particle production in the world-volume of the probe brane. For the particular case of a scalar field living in the world-volume of a probe $D6$-brane, we show that particle production may occur in a regime of parametric resonance similar to the one behind the phenomenon of preheating after inflation \cite{preheating,Kofman}, leading to an exponential growth of the particle number in the probe's world-volume.\footnote{Discussions of preheating in brane systems other than branonium can be found in \cite{otherbranepreheating}.} We determine how the orbital parameters of the system constrain this mechanism and analyze the possible properties of the resonantly produced particles. Also, we study how the probe's trajectory is modified by compactifying three of the dimensions parallel to the branes, so that a 4-dimensional spacetime with a flat Friedmann-Robertson-Walker geometry is obtained. In this case, we analyze the effects of Hubble expansion on the orbital parameters and how it modifies the parametric resonance regime. Other instability sources such as radiation into bulk modes are discussed as well. We also consider the effects of transverse space compactification on the probe's motion in the case of a compact 3-torus and analyze the associated production of angular momentum. Finally, we explore some of the cosmological applications of the resonant particle production mechanism and relate them to possible final configurations of the system, including stable states or brane-antibrane annihilation.

In the next section, we begin by introducing some of the general properties of $Dp-\overline{Dp}$ branonia, with particular emphasis on the $p=6$ case, and discuss the effects of Hubble expansion on the orbital parameters. In Section III, we study the evolution of scalar particle-modes confined to the probe brane and analyze the properties of the associated parametric resonance mechanism in both the non-expanding and expanding universe cases. We distinguish between the production of massless and massive particles and determine the significance of energy damping into these modes on the probe's motion. We also consider, in this section, the effects of radiation into bulk modes on the resonant production of brane-bound particles. In Section IV, we analyze how compactifying the directions transverse to the branes modifies the probe's trajectory and dynamically generates the required angular momentum. Possible final states of the system are described in section V, where we also analyze the relevance of the particle production mechanism to cosmology in some particular scenarios. We summarize the main results and conclusions of this work in section VI.


\section{Evolution of branonium in an expanding universe}

Branonium corresponds, in its simplest form proposed in \cite{Burgess}, to a system whose components are a fixed stack of $N$ parallel $p$-branes and a probe $p$-brane, parallel to the stack, whose gravitational backreaction is neglected so that it has no influence on the geometry of the background spacetime. Let the coordinates $x^{\mu}$, $\mu=0,\cdots,p$, correspond to the dimensions parallel to the stack and the coordinates $y^m$, $m=p+1,\cdots,D-p$, correspond to those transverse to the branes. Then, the general solution of the supergravity equations of motion in $D$ dimensions for $N$ parallel $p$-branes is given, in the Einstein frame, by \cite{Stelle}: 
\beqa \label{eq1}
ds^2&=&h^{-\tilde{\gamma}}dx_p^2+h^{\gamma}dy^2~,\nonumber\\
e^{\Phi}&=&h^{\kappa}~,\qquad C_{01\cdots p}=\zeta(1-h^{-1})~,
\eeqa
which give the $D$-dimensional metric, the dilaton $\Phi$ and the non-vanishing components of the bulk $(p+1)$-form field $C_{[p+1]}$ which couples to the $(p+1)$-dimensional world-volume of the stack of branes.  The harmonic function $h(r)$ is a function of the radial coordinate in transverse space, $r^2=\delta_{mn}y^my^n$, and has the form:
\beq \label{eq2}
h(r)=1+{Q_p\over r^{\tilde{d}}}~,
\eeq
where $\tilde{d}=D-d-2$ and the factor of $1$ ensures that the solution is asymptotically the $D$-dimensional Minkowski space. For the cases we will be interested in studying, the transverse space is at least 2-dimensional, so that $\tilde{d}>0$. The stack of branes is, thus, located at the origin in the transverse space. The exponents $\gamma$, $\tilde{\gamma}$, $\kappa$ and the constant $\zeta$ that characterize the solution are given by:
\beqa \label{eq3}
\gamma&=&{4d\over(D-2)\Delta}~,\qquad \tilde\gamma={4\tilde{d}\over(D-2)\Delta}~,\nonumber\\
\kappa&=&{2\alpha\over\Delta}~,\qquad\qquad\ \ \ \zeta={2\over\sqrt{\Delta}}~,
\eeqa
where $d=p+1$ and
\beq \label{eq4}
\Delta=\alpha_n^2+{2d\tilde{d}\over D-2}~.
\eeq

The constant $\alpha_n$ defines the coupling of the $n=(p+2)$-form field strength to the dilaton in the supergravity action. We will be interested in the particular case of $Dp$-branes in Type II supergravities, which arise as the low energy limit of Type II string theories, so that the $n$-form field strength corresponds to a closed string state in the Ramond-Ramond sector, and $\alpha_n=\alpha_R=2(D-2n)/(D-2)$. Taking $D=10$, we have $\alpha_R={1\over2}(3-p)$, $\Delta=4$, $\kappa={1\over4}(3-p)$, $\tilde{d}=7-p$, $\tilde{\gamma}={1\over8}(7-p)$ and $\gamma={1\over8}(p+1)$. It is clear that the condition $\tilde{d}>0$ implies $p<7$. The constant $Q_p$ is related to the charge carried by the stack of $N$ branes, and for the case of $Dp$-branes we have
\beq \label{eq5}
Q_p=c_pg_sNl_s^{\tilde{d}}~,
\eeq
where $c_p=(2\sqrt{\pi})^{5-p}\Gamma\big({7-p\over2}\big)$, $l_s$ is the string length and $g_s$ is the string coupling constant. This solution holds in the limit of validity of string perturbation theory and for small curvature, which allows a low energy description of the system in terms of supergravity fields. This approximation is valid as long as the local string coupling, $g_se^{\Phi}$, is everywhere small. Also, we need to require the radial coordinate $r$, which will give the distance between the probe and the stack branes, to be large compared to the string length $l_s$, so that mediation by supergravity bulk fields (graviton, dilaton and RR-forms) is the dominant source of brane interactions.

Before analyzing the motion of the probe brane, we need to consider possible compactifications of the 10 dimensional spacetime that yield our 4-dimensional world at low energies. Compactification of dimensions parallel and transverse to the world volume of the branes involve different procedures. In the case of dimensions parallel to the branes, it suffices to consider the Kaluza-Klein \textit{ansatz}. If we start with the $D$-dimensional spacetime Eq. (\ref{eq1}) and compactify along a periodic coordinate $z\simeq z+2\pi R$ with radius $R$, so that the branes are wrapped around this dimension, we obtain a $D'=(D-1)$-dimensional spacetime with metric given by \cite{Stelle}:
\beqa \label{eq6}
ds_D^2&=&e^{2\hat{a}\varphi}ds_{D'}^2+e^{2\hat{b}\varphi}(dz+{\cal B}_{\mu}dx^{\mu})^2 ~,\nonumber\\
ds_{D'}^2&=&h^{-\tilde{\gamma'}}dx_{p'}^2+h^{\gamma'}dy^2~,\nonumber\\
e^{\varphi}&=&h^{\rho}~,
\eeqa 
where
\beqa \label{eq7}
\hat{a}&=&{1\over2(D'-1)(D'-2)}~,\qquad \hat{b}=-(D'-2)\hat{a}~,\nonumber\\
\rho&=&{\tilde{\gamma}\over2(d'-2)\hat{a}}~,\nonumber\\
\gamma'&=&\gamma-2\hat{\rho}~,\qquad \tilde{\gamma}'=\tilde{\gamma}+2\hat{a}\rho~.
\eeqa

The dilaton field $\Phi$ is not altered by this compactification, while the $(n-1)$-form RR field can be decomposed as follows:
\beq \label{eq8}
C_{[n-1]}=B_{[n-1]}+B_{[n-2]}\wedge dz~.
\eeq

We take the following truncation of these fields:
\beq \label{eq9}
B_{[n-1]}=0~,\qquad (B_{[n-2]})_{01\cdots p-1}=\zeta(1-h^{-1})~.
\eeq

This truncation leads to a $(p-1)$-brane supergravity solution in $D-1$ dimensions, characterized by the same harmonic function of the transverse space radial coordinate $h(r)$. Applying this \textit{toroidal compactification} procedure $j$ times we obtain $(p-j)$-brane solutions living in a $(D-j)$-dimensional spacetime, with the number of transverse dimensions remaining the same.

The process of compactifying the dimensions transverse to the branes is not as straightforward due to the lack of translational invariance which results from the presence of the branes. In particular, to make a given transverse coordinate $z'$ periodic, so that a similar Kaluza-Klein mechanism can be applied, one needs to include the appropriate ``image branes" of the source branes which will ensure the invariance of the configuration under $z'\rightarrow z'+2\pi R'$, $R'$ being the radius of the compact dimension. For example, if a source brane is located at $z'=0$, an image brane has to be considered at $z'=2\pi R'$. This implies a generalization of the harmonic function in Eq. (\ref{eq2}) to include branes located at different points in transverse space \cite{Stelle}:
\beq \label{eq10}
h(\mathbf{y})=1+\sum_i{Q_p\over |\mathbf{y}-\mathbf{y_i}|^{\tilde{d}}}~,
\eeq
where the vectors $\mathbf{y_i}$ denote the positions of the different image branes and $|\mathbf{y}-\mathbf{y_i}|^2=\delta_{mn}(y^m-y_i^m)(y^n-y_i^n)$.  

These two compactification schemes allow us to reduce a 10-dimensional solution of the supergravity field equations to a 4-dimensional spacetime which resembles our world. We will be interested in studying a particular case of the general branonium configuration described so far where we have a source stack of D6-branes and also a probe $D6$ brane. As was shown in \cite{Burgess}, when the probe carries an antibrane charge, i.e. opposite to the charge of the source branes, the system yields closed elliptical orbits which can be solved analytically, while for other values of $p<7$ the orbits fail to close. This feature has, as we will describe later on, important consequences for particle production in the world-volume of the probe.

Compactifying 3 of the dimensions parallel to the branes, we obtain effective D3-branes moving in a 3-dimensional transverse space, the shape of the orbits not being affected by this toroidal compactification scheme as mentioned before. If the transverse dimensions are finite but their typical size is much larger than that of the compact parallel dimensions and the interbrane distance, we may consider them to be infinite and neglect the brane images described before. Although this would correspond to a 7-dimensional spacetime, from the point of view of fields confined to the world-volume of the source and probe branes it is effectively 4-dimensional. The dynamics in the transverse space will, nevertheless, affect the 4-dimensional dynamics of the fields.

In order to make the system more similar to the observable universe, we consider a modification of the solution Eq. (1) so that, after compactification of 3 of the parallel dimensions, the 4-dimensional world-volume of the branes has a flat Friedmann-Robertson-Walker (FRW) geometry. This is a particular case of the the system studied in \cite{Lukas}, where the moduli governing the size of both parallel and transverse dimensions were allowed to vary. In this work, we will assume that some dynamical mechanism at the string scale (see, e.g. \cite{DeWolfe}) or at some other high-energy scale (such as SUSY breaking) fixes all moduli except the scale factor associated with the 3 non-compact dimensions parallel to the branes. Hence, in the Einstein frame, the complete 7-dimensional line element is given by:
\beq \label{eq11}
ds_7^2=h^{-\tilde{\gamma}_4}[-dt^2+a^2\delta_{ij}dx^idx^j]+h^{\gamma_4}\delta_{mn}dy^mdy^n~,
\eeq 
where $a\equiv a(t)$, $i,j=1,2,3$ and $m,n=1,2,3$. The harmonic function is in this case $h(r)=1+Q_6/r$. The exponents $\gamma_4$ and $\tilde{\gamma}_4$ can be obtained using Eq. (\ref{eq3}) and applying Eq. (\ref{eq7}) for the compactification on a 3-torus, but their precise values will not be necessary in the subsequent calculations.

The motion of the probe brane is described by its action, which is divided in two parts. The Born-Infeld contribution, which concerns the geometry of the world volume and possible gauge fields living on it, is in the string frame given by, for a generic $p$-brane probe \cite{Burgess}:
\beq \label{eq12}
S_{BI}=-T_p\int d^{p+1}\xi\ e^{-\Phi}\sqrt{-det(\hat{\gamma}_{\mu\nu}+2l_s^2\cal{F}_{\mu\nu})}~,
\eeq
where $\xi^{\mu}$ are world-volume coordinates, $T_p$ gives the $p$-brane tension and $\hat{\gamma}_{\mu\nu}=\hat{g}_{MN}\partial_{\mu}x^M\partial_{\nu}x^N$ is the induced metric on the probe. The hatted metric tensor is defined in the string frame, and is related to the Einstein frame one via $\hat{g}_{\mu\nu}=e^{\lambda\Phi}g_{\mu\nu}$ with $\lambda=4/(D-2)$ in the general case. The antisymmetric field strength tensor $\cal{F}_{\mu\nu}$ refers to gauge fields describing open string modes with endpoints on the probe (in the case of Dirichlet branes), but in what follows we will set these fields to zero\footnote{This means that we will consider the particular scenario where the classical background of these gauge fields vanishes. The role of gauge fields was considered in \cite{Takahashi}.}. The Wess-Zumino part of the action describes the minimal coupling of the Ramond-Ramond $(p+1)$-form to the $(p+1)$-world volume of the probe brane and can be written as \cite{Burgess}:
\beq \label{eq13}
S_{WZ}=-qT_p\int C_{[p+1]}~.
\eeq 

The constant $q$ gives the sign of the charge of the probe brane, which being a BPS saturated state is equal or opposite to the brane tension $T_p$, so that $q=1$ corresponds to a brane and $q=-1$ corresponds to an antibrane \cite{Burgess}. 

If we now consider the particular case of D6 branes in 10 dimensions, compactifying 3 of the parallel dimensions and including the scale factor describing the expansion (or contraction) of the 3 remaining non-compact dimensions along the brane, and choose the world-volume coordinates such that $\xi^{\mu}=x^{\mu}$, we arrive at the following expression for the total action of the probe brane:
\beq \label{eq14}
S_4=-T_6V_3\int d^4x\ a^3(t)\bigg[{1\over h}\sqrt{1-hv^2}+q\bigg(1-{1\over h}\bigg)\bigg]~,
\eeq
where $V_3$ gives the volume of the compact 3-torus around which the D6-branes are wrapped to give the effective 4-dimensional space. We have assumed that the transverse space coordinates are exclusively time-dependent, $y^m=y^m(t)$ and defined the velocity of the probe as $v^2=\delta_{mn}\dot{y}^m\dot{y}^n$, with dots denoting time derivatives. As mentioned before, the bound orbits of the probe brane can be exactly solved for the case where the scale factor is constant and it was shown in \cite{Burgess} that they correspond to closed ellipses. We now wish to analyze the effects of the varying scale factor on these orbits. In this case, a full analytical solution is hard to obtain and it is simpler to consider the evolution of the system when the probe brane moves at larges distances from the stack with small velocities. These two conditions can be expressed as $hv^2\ll1$. One must recall that we considered initially that the transverse dimensions can be considered infinite in extent. Hence, our analysis will hold as long as the radial coordinate is large compared to the typical length scale of the harmonic function, given by $Q_6$ (assumed larger than the string scale and the compactification scale), but still much smaller than the typical size of the transverse dimensions (assumed much larger than $Q_6$). With this hierarchy in mind, we may expand the probe brane's action to obtain, to lowest order, 
\beq \label{eq15}
S_4\approx-T_6V_3\int d^4x\ a^3(t)\bigg[{1\over h}\bigg(1-{1\over2}hv^2-q\bigg)+q\bigg]~.
\eeq

Our subsequent analysis will not be affected by the constant term and we may drop it. For a probe antibrane,
\beq \label{eq16}
S_4\approx T_6V_3\int d^4x\ a^3(t)\bigg[{1\over2}\sum_i\dot{y}_i^2-{2\over h} \bigg]~.
\eeq

The rotational symmetry of the problem implies that the probe's trajectory will be confined to a plane, which we may choose to be $y^3=0$. Then, we may define the canonically normalized complex scalar field
\beq \label{eq17}
\phi\equiv\sqrt{{T_6V_3\over2}}(y^1+iy^2)~,
\eeq
so that its action is, in the non-relativistic and large distance limit,
\beq \label{eq18}
S_{\phi}=\int d^4x\ a^3(t)\big(-g^{\mu\nu}\partial_{\mu}\phi\partial_{\nu}\phi^{*}-V(\phi)\big)~,
\eeq
where $g_{\mu\nu}$ is the 4-dimensional flat FRW metric of the spacetime where the field is defined and the potential is given by:
\beq \label{eq19}
V(\phi)=2T_6V_3h^{-1}\approx2T_6V_3\bigg(1-\sqrt{{T_6V_3\over2}}{Q_6\over|\phi|}\bigg)~,
\eeq
so that it only depends on $\rho=|\phi|$. Note that in writing the action in this form one must take into account that the spatial derivatives of the field, $\partial_i\phi$, $i=1,2,3$, are assumed to vanish, i.e. the probe brane should remain parallel to the source branes during its motion. The analysis of the classical motion of the probe brane is in this way reexpressed as a classical field theory problem of a complex scalar field in the background of a flat FRW spacetime. Before deriving the equations of motion for the field $\phi$, it is useful to consider some of the symmetries of the action. First, the rotational symmetry of the problem in the plane transverse to the branes is associated with a conserved angular momentum. In terms of the field theory approach, this corresponds to the global U(1) invariance of the action under $\phi\rightarrow e^{i\alpha}\phi$, where $\alpha$ is a constant. The associated Noether current induces a conserved charge which gives, in the quantum theory, the particle number operator associated with the field. Writing the field as $\phi=\rho e^{i\theta}$ and defining the comoving angular momentum as $l\equiv\rho^2\dot\theta$, the comoving particle number density is given by:
\beq \label{eq20}
n=i(\dot\phi^*\phi-\dot\phi\phi^*)=2\rho^2\dot\theta=2l~,
\eeq
while the total particle number, which corresponds to the conserved Noether charge, is given by $N=a^3n$. The energy density and pressure of the field can be obtained from its energy-momentum tensor, yielding :
\beqa \label{eq21}
\epsilon&\equiv& T_{00}=|\dot\phi|^2+V(\phi)=\dot\rho^2+{l^2\over\rho^2}+V(\rho)~,\nonumber\\
p&\equiv& {T_{ii}\over a^2}=|\dot\phi|^2-V(\phi)=\dot\rho^2+{l^2\over\rho^2}-V(\rho)~.
\eeqa

The covariant conservation of the energy-momentum tensor can then be expressed as $dE=-pdV$, where the energy of the field is defined as $E\equiv a^3\epsilon$ and $V=a^3$ is the volume of the expanding (or contracting) flat FRW spacetime. Defining the Hubble parameter $H\equiv\dot a/a$, we can write this in the form:
\beq \label{eq22}
{d\epsilon\over dt}=-3H(\epsilon+p)~.
\eeq

From Eqs. (\ref{eq21}) and (\ref{eq22}) we conclude that ${d\epsilon\over dt}\leq 0$, the equality holding only for $H=0$. Thus, in an expanding universe ($H>0$), the system evolves with a strictly decreasing energy density. The angular momentum evolves according to:
\beq \label{eq23}
{dl\over dt}+3Hl=0~,
\eeq
which corresponds to conservation of the total particle number $N$. Hence, in an expanding universe, the absolute value of the angular momentum will necessarily decrease, vanishing asymptotically.

We, thus, conclude from this simple analysis of conservation laws that the expansion of the universe will make the system reduce both its energy density and its angular momentum. To analyze the details of this evolution, we need to determine the equations of motion for the field $\phi$. Varying the action Eq. (\ref{eq18}) with respect to $\phi^*$, and taking this variation to vanish, we obtain:
\beq \label{eq24}
\ddot\phi+3H\dot\phi+{\partial V(|\phi|)\over\partial\phi^*}=0~.
\eeq
In terms of the transverse space coordinates $y^m$, $m=1,2,3$, this can be written as:
\beq \label{eq25}
\ddot y^m+3H\dot y^m+2Q_6{y^m\over r^3}=0~.
\eeq

This implies that the polar variables $\rho$ and $\theta$ satisfy:
\beqa \label{eq26}
&\ddot\rho+3H\dot\rho-{l^2\over\rho^3}+{1\over2}{\sigma\over\rho^2}=0~,\nonumber\\
&\ddot\theta+2{\dot\rho\over\rho}\dot\theta+3H\dot\theta=0~,
\eeqa
where we have defined the constant 
\beq \label{eq27}
\sigma\equiv4Q_6\bigg({T_6V_3\over2}\bigg)^{3/2}~,
\eeq
so that, apart from constant factors, the large distances potential is $V(\rho)=-{\sigma\over\rho}$. It is easy to check that the equation for $\theta$ simply gives the evolution of the angular momentum that we obtained previously in Eq. (\ref{eq23}).

The evolution of the radial field $\rho$ is determined by the form of the Hubble parameter, given by the energy density content of the universe via the Friedmann equation in the usual way. This may include not only the energy density of the field $\phi$ but also all other matter, radiation or vacuum energy components. The possibility that this scalar field may drive (slow-roll) inflation if, in the early universe, it dominates the energy density has been analyzed in \cite{Quevedo}, for the case where the probe brane moves in a linear trajectory, i.e. $l=0$. It was shown that the interbrane potential arising from Type II supergravity/string theories is not flat enough to produce the required number of e-foldings, exhibiting an ``$\eta$-problem". This is intrinsically related to the assumption that $r$ is much smaller than the size of the transverse dimensions. 

An inflationary period driven by the angular field variable $\theta$ was proposed in \cite{Burgess}, where it is argued that, despite having a flat potential, this field may provide the constant energy density necessary for inflation if the probe brane moves in a circular orbit with $\dot\rho=0$. However, we need to take into account the fact that, during inflation, the angular momentum decays exponentially as $l\propto e^{-3Ht}$, from Eq. (\ref{eq23}). The system will hence quickly tend to the $l=0$ case, where sufficient inflation is difficult to obtain. Furthermore, the decay of the angular momentum will necessarily alter the evolution of $\rho$ and the value of $H$, so that a slow-roll inflationary mechanism with the probe brane moving in a circular orbit would be hard to construct\footnote{Some work has been done recently in the context of $D$-branes moving in warped throats \cite{Kehagias}. In particular, the important role of the angular variables in providing accelerated periods of expansion was discussed in \cite{Easson}.}. 

Although we do not wish to completely discard such inflationary mechanisms, we will from now on assume that some other field is responsible for inflation and analyze how the compactified $D6-\overline{D6}$ branonium system evolves in the post-inflationary eras.

Before analyzing in detail the effects of the expansion on the trajectory of the probe, let us recall the general bound orbits of the system in a non-expanding universe. In this case, the system reduces to the well-known problem of a particle in a central $1/\rho$ attractive potential, admitting closed orbit solutions of the form:
\beq \label{eq28}
\rho(\theta)={R(1-e^2)\over 1+e\cos\theta}~.
\eeq
These are closed ellipses with eccentricity and semi-major axis given by:
\beqa \label{eq29}
e&=&\bigg(1+{4\epsilon l^2\over\sigma^2}\bigg)^{1\over2}~,\nonumber\\
R&=&-{\sigma\over2\epsilon}~.
\eeqa

The properties of such orbits may be obtained by studying the effective potential $V_{eff}(\rho)=l^2/\rho^2+V(\rho)$. This potential has a minimum at $-(\sigma/2l)^2$, tends to $+\infty$ at the origin and to zero at large distances. The condition $\dot\rho\geq 0$ then implies that orbits with $-(\sigma/2l)^2<\epsilon<0$ will be bounded, with $0<e<1$. In this case, $\rho$ will oscillate between its minimum and maximum values, $\rho_{min}=R(1-e)$ and $\rho_{max}=R(1+e)$, which can be obtained from Eq. (\ref{eq28}) by setting $\theta=0$ and $\theta=\pi$, respectively. At the minimum of $V_{eff}(\rho)$, the orbits will be circular with $\rho=\rho_c=2l^2/\sigma$, and $e=0$. For $\epsilon\geq0$, the orbits will only be bounded from below with e.g. $\rho\geq l^2/\sigma$ for $\epsilon=0$, which corresponds to an $e=1$ parabolic orbit. The linear trajectory, with $l=0$, is a particular case of the latter with no bounds on the value of $\rho$ except for the trivial $\rho\geq0$. For $\epsilon>0$, the orbits are hyperbolic.

Consider now the expanding case with the scale factor evolving as $a(t)\propto t^{\alpha}$. This power law behavior is typical of the post-inflationary stages of the universe's evolution, where one may consider a single fluid to give the dominant contribution to the total energy density. For example, in the radiation era we have $\alpha=1/2$ and in the matter era $\alpha=2/3$. We can also take $\alpha=2/3$ at the end of inflation, when the oscillations of the inflaton about the minimum of its potential, with an equation of state corresponding to that of non-relativistic matter, dominate the energy density. This model only holds away from the transitions between these periods as in these cases at least two of the components give similar contributions to the energy density.

We will consider the evolution of the branonium system for arbitrary $\alpha$, starting at some instant $t_0$ when the scale factor has a value $a_0\equiv a(t_0)$. The Hubble parameter is then of the form $H={\alpha\over t}$, which implies from Eq. (\ref{eq23}) that the angular momentum evolves according to:
\beq \label{eq30}
l=l_0\bigg({t\over t_0}\bigg)^{-3\alpha}~,
\eeq
with $l_0$ being its initial value. It is then clear that $l\rightarrow 0$ as $t\rightarrow +\infty$, as we have mentioned earlier. The equation for the radial field $\rho$ can now be written as:
\beq \label{eq31}
\ddot\rho+{3\alpha\over t}\dot\rho-{l_0^2\over\rho^3}\bigg({t\over t_0}\bigg)^{-6\alpha}+{1\over2}{\sigma\over\rho^2}=0~.
\eeq

Let us start by analyzing how the system evolves when placed initially in a would-be circular orbit, so that
\beq \label{eq32}
\rho(t_0)={2l_0^2\over\sigma}\equiv\rho_{c0}~,
\eeq
as $\ddot\rho(t_0)=\dot\rho(t_0)=0$. It is easy to see that $\rho$ cannot remain constant for $t>t_0$ due to the decay of the angular momentum. We may, however, look for solutions where the condition for circular orbits is maintained during the motion of the probe brane, i.e. 
\beq \label{eq33}
\rho(t)={2l^2(t)\over\sigma}=\rho_{c0}\bigg({t\over t_0}\bigg)^{-6\alpha}\equiv\rho_{c}(t)~. 
\eeq

Substituting into Eq. (\ref{eq31}) we obtain
\beq \label{eq34}
{\rho_{c0}\over t_0^2}\bigg({t\over t_0}\bigg)^{-6\alpha-2}6\alpha(3\alpha+1)=0~,
\eeq
which is satisfied only in the non-expanding case, $\alpha=0$, and for $\alpha=-1/3$. As we are interested in cases where $\alpha>0$, we conclude that $\rho_c(t)$ is not an exact solution of the equations of motion. It is, however, an approximate solution at late times and, as we will check later, gives the global evolution of the system in an expanding universe, so that it is useful to consider some of its properties. In particular, the angular frequency evolves as:
\beq \label{eq35}
\dot\theta_c={l\over\rho_c^2}\propto\bigg({t\over t_0}\bigg)^{9\alpha}~,
\eeq 
so that the orbital period decreases as $T_c(t)\propto(t/t_0)^{-9\alpha}$. Such a solution eventually fails to satisfy the non-relativistic and large distance approximation as the radius of the orbit decreases and its angular velocity grows. In particular, the parameter $hv^2$ which controls this approximation grows, for $t\gg t_0$, as
\beq \label{eq36}
hv^2\rightarrow{1\over2}(T_6V_3)^4\bigg({Q_6\over l_0}\bigg)^4\bigg({t\over t_0}\bigg)^{12\alpha}~.
\eeq
Hence, given the values of $T_6$ and $V_3$, it is the ratio $Q_6/l_0$ that controls how long the approximations remain valid. 

Consider now trajectories with a non-vanishing initial eccentricity. In an expanding universe, the effective potential becomes time-dependent, although its asymptotic properties at the origin and at infinity remain the same. As the angular momentum redshifts away, the minimum of the effective potential decreases, the same happening with the value of $\rho$ at this minimum, according to the circular solution defined in Eq. (\ref{eq33}). As discussed earlier, the energy density of the field always decreases during the motion of the probe brane in an expanding universe. Hence, if the system is initially in one of the closed orbits with $\epsilon<0$, the latter condition will be satisfied for all $t>t_0$. This means that, at all times, the probe brane will be in one of the closed elliptical orbits defined earlier. From Eq. (\ref{eq29}), we also conclude that the semi-major axis of the orbit will decrease as the energy density evolves to more negative values. The eccentricity of the orbits may, however, not remain constant, as it depends on the variation of both the energy density and the angular momentum. In fact, using Eqs. (\ref{eq22}) and (\ref{eq23}), one can show that the eccentricity varies according to
\beq \label{eq37}
{de^2\over dt}=-12H{l^2\over\sigma^2}(3\epsilon+p)=-24H{l^2\over\sigma^2}(2T+V)~.
\eeq
where we defined the kinetic energy of the field as $T=\dot\rho^2+{l^2\over\rho^2}$, so that $\epsilon=T+V$ and $p=T-V$.

Thus, in an expanding universe, for finite $l_0$, one expects the eccentricity of the orbit to vary unless $T=-V/2$. This last condition is satisfied on average in the non-expanding case, corresponding to the Virial Theorem. It does not necessarily hold, however, in an expanding universe, with the semi-major axis of the orbits decaying as discussed above. It is nevertheless clear that, at late times, as $H$ and $l$ decrease, the variation of the eccentricity of the orbits should be smaller. The sign of $2T+V$ is also not definite, so that the eccentricity may either increase or decrease during the motion of the probe brane.

From Eqs. (\ref{eq29}) and (\ref{eq33}), we obtain:
\beq \label{eq38}
R(t)={\rho_c(t)\over 1-e^2(t)}~.
\eeq
Hence, the variation of the semi-major axis follows the decrease of the circular solution obtained earlier, being also affected by the variation of the eccentricity. 

In order to have a better understanding of the evolution of the probe brane's motion in an expanding universe, we have solved the equations of motion numerically. Measuring all quantities in terms of the string length, i.e. setting $l_s=1$, we choose, as an example, the values $V_3=(T_6)^{-1}=(2\pi)^{-6}$ and $Q_6=100$\footnote{For the value of $Q_6$ we follow the example analyzed in \cite{Lukas}.}. We set the initial conditions at $t=t_0$ to be those of a non-expanding elliptical orbit with angular momentum $l_0$ and eccentricity $e_0$, which we take to be the only free orbital parameters. The field is initially at its maximum value $\rho_0={\rho_{c0}\over1-e_0}$, with $\dot\rho_0=0$ and $\theta=\pi$. After determining the numerical solution for $\rho(t)$, we computed the corresponding energy density and eccentricity evolution, according to Eqs. (\ref{eq21}) and (\ref{eq29}), respectively. We used the numerical solution for the eccentricity to compute $R(t)$, according to Eq. (\ref{eq38}), and also the maximum and minimum values of $\rho(t)$ at each orbit, given by:
\beqa \label{eq39}
\rho_{min}(t)&=&R(t)(1-e(t))~,\nonumber\\
\rho_{max}(t)&=&R(t)(1+e(t))~.
\eeqa
We have also determined the evolution of the angular field variable $\theta$ using $\dot\theta(t)=l(t)/\rho(t)^2$.

In Figure 1 we have plotted the results obtained for the values $l_0=2000$\footnote{The non-relativistic and large distance approximation holds for this value of initial angular momentum.} and $e=0.2$ in a matter-dominated universe, $\alpha=2/3$, with $t_0=10^8$ \footnote{Notice that although this value of $t_0$ seems quite large, one must take into account that it is measured in units of the string time, which is of order $10^{-43}$ seconds if the string scale is of the order of the Planck scale. Thus, we are considering the probe brane to be moving quite early in the history of the universe.}.

\begin{figure}[htbp]
	\centering
		\includegraphics[scale=0.7]{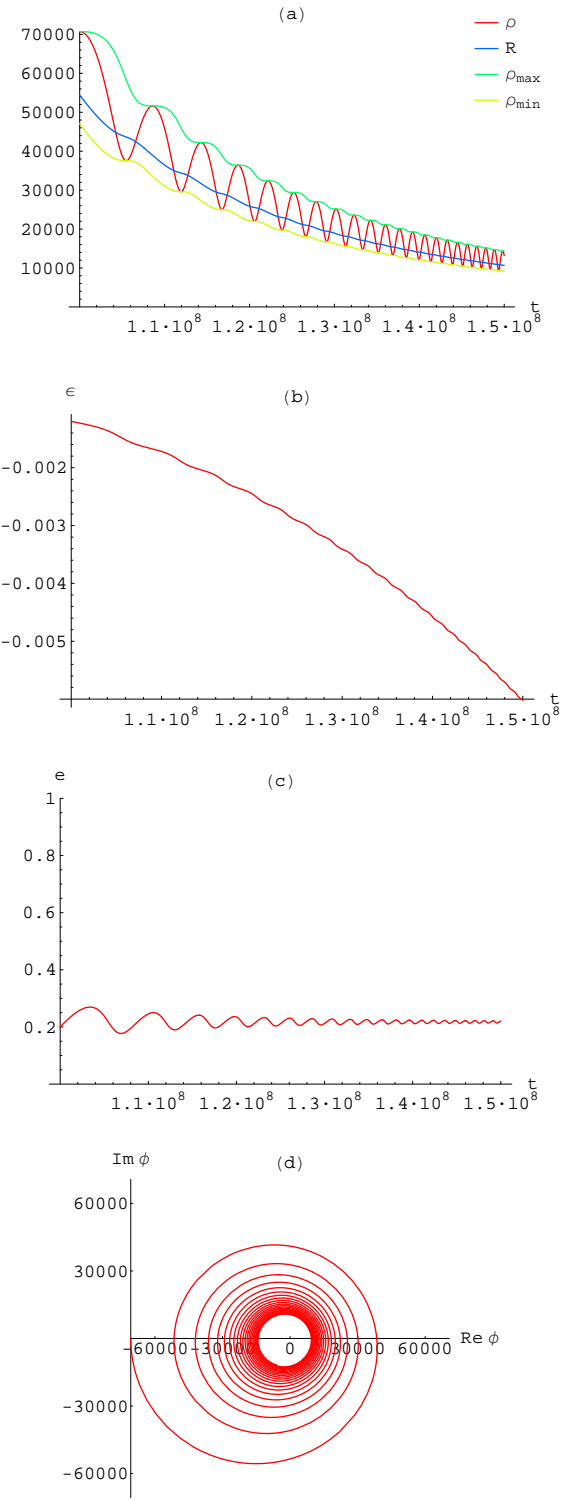}
	\label{fig:figure_1}
	\caption{Numerical results obtained for $l_0=2000$ and $e=0.2$ in a matter-dominated universe, $\alpha=2/3$, with $t_0=10^8$. The plots show (a) the radial field and associated quantities, (b) the energy density, (c) the eccentricity and (d) the motion of the probe brane in the complex plane of the field $\phi$.}
\end{figure}

Observing the plots shown in this figure, we conclude that, as expected, the probe brane evolves continually through elliptical orbits of decreasing semi-major axis, so that they fail to close. The radial field then oscillates between minimum and maximum values which decrease in time. The global decrease follows that of the circular solution, Eq. (\ref{eq33}), but exhibits an oscillating behavior associated with the variation of the energy density. Although the latter strictly decreases, as expected, it also oscillates with decreasing period and amplitude, so that at later stages its evolution tends to be smooth. Consequently, one observes oscillations in the evolution of the eccentricity of the orbit. In the example shown in Figure 1, these oscillations have a small amplitude. However, at earlier times, when the effects of Hubble expansion are more significant, numerical simulations show that this amplitude can be quite large. All simulations show that $e(t)$ starts increasing and, as the period and amplitude of the oscillations decay, it tends to a constant value. 

A constant eccentricity at late times implies that, in this limit, the energy density should, on average, vary as
\beq \label{eq40}
\vev{\epsilon(t)}\propto l^{-2}(t)\propto\bigg({t\over t_0}\bigg)^{6\alpha}~,
\eeq
tending smoothly to more negative values as observed in Figure 1. Then, from Eq. (\ref{eq22}),
\beq \label{eq41}
\vev{{d\epsilon\over dt}}=-3H\vev{\epsilon+p}=6H\vev{\epsilon}~,
\eeq
so that $\vev{3\epsilon+p}=0$ or, equivalently, $\vev{2T+V}=0$. Thus, although at early times the effect of the expansion makes the eccentricity vary significantly, at late times the system \textit{virializes} and starts evolving smoothly between orbits of constant eccentricity and decaying radius. This result will be useful later for studying particle production in the probe brane at late times.

Another interesting consequence of the expansion of the universe is a rotation of the axis of the elliptical orbits with time, although this is a small effect in the example shown in Figure 1. This is simply a result of the growing eccentricity of the orbits which deviates the maximum and minimum values of $\rho$ from $\theta=\pi$ and $\theta=0$, respectively. The orbital axis will, however, stabilize at late times, as $e(t)$ tends to a constant value.

Numerically, one also observes a decrease in the amplitude of the oscillations of $\rho(t)$. This is given by
\beq \label{eq42}
\Delta\rho(t)\equiv\rho_{max}(t)-\rho_{min}(t)=R(t)e(t)~,
\eeq
so that the observed decay of the amplitude is mainly due to the decrease of $R(t)$ as discussed above.

The asymptotic value of the eccentricity depends not only on the initial time $t_0$ at which the probe starts its motion but also on the initial values of the orbital angular momentum and eccentricity. Numerical simulations show that larger values of $l_0$ lead to a more pronounced growth of the eccentricity, in agreement with Eq. (\ref{eq37}). The dependence on $e_0$ is less trivial and numerically one finds a larger eccentricity variation for $e_0$ close to 0 and to $0.9$. This is, however, very small for $e_0$ close to 1, which is expected, as the energy density of the field cannot increase to produce hyperbolic orbits with $e>1$ if initially $\epsilon<0$. The eccentricity always grows if initially $\dot\rho=0$, so that $2T+V<0$ and ${d\epsilon\over dt}>0$, as most of the eccentricity variation occurs initially when Hubble expansion is more significant. The opposite behavior should be observed if initially $2T+V>0$. 

Although the example we have shown refers to a matter-dominated universe, $\alpha=2/3$, our qualitative discussion and analytical results hold for all $\alpha>0$, in particular to the $\alpha=1/2$ radiation-dominated universe. 

To summarize the results of this section, we conclude that, in an expanding universe, the probe brane follows elliptical orbits with decreasing radius and increasing frequency and whose eccentricity exhibits an oscillating behavior and asymptotically tends to a constant value larger than the initial one. If the motion of the probe brane begins at late times with a small angular momentum, the latter effect is negligible.

The probe brane will asymptotically collide with the branes in the stack, as $\rho\rightarrow0$, but our approximations will break down before this happens, so that we expect the motion of the probe brane to deviate significantly from our previous results as its angular velocity grows and it spirals towards the central stack. The collision will, however, occur if nothing else prevents the probe from losing energy and angular momentum as the universe expands, and may lead to the annihilation of the probe antibrane with one of the source branes. For now, we will use the results obtained in this section to study the production of particles in the probe brane and we will return to this issue in section V of this work.


\section{Particle Production}

In the original discussion of the branonium system \cite{Burgess}, it was argued that the fields confined to the probe brane, such as the gauge fields mentioned earlier, become time-dependent due to the motion of the probe through the background spacetime created by the central stack. Consequently, an observer bound to the probe interprets this variation as corresponding to the production of particles associated with these fields. Such radiation of energy into brane particle-modes arises only if the distance between the probe and the stack varies in time, thus creating a time-dependent background from the point of view of brane-bound observers. Although the power radiated into these modes was estimated in \cite{Burgess}, many aspects of this particle production mechanism remain unclear. In this section, we will analyze this mechanism in more detail, revealing some new properties of particle production in branonium, and discuss how the expansion of the universe modifies this process.

The fields confined to the brane arise, from the string theory point of view, as states associated with open strings whose endpoints are attached to the brane, as discussed earlier. For a single $p$-brane, which is a $1/2$ BPS supersymmetric state, these include bosonic fields such as gauge bosons and scalar fields, as well as their fermionic superpartners. If the probe corresponds itself to a stack of $M$ parallel $p$-branes, other degrees of freedom associated with a $U(M)$ vector supermultiplet may arise \cite{Becker}. Here, we will study the simplest case of a real scalar field $\eta$ confined to the probe brane and follow an effective field theory approach. 

For the $D6-\overline{D6}$ system described earlier, we write the effective action for the scalar field $\eta$ in the form:
\beq \label{eq43}
S_{\eta}=-T_6\int d^7\xi\  e^{-\Phi}\sqrt{-\hat{\gamma}}\bigg(-{1\over2}\hat\gamma^{\mu\nu}\partial_{\mu}\eta\partial_{\nu}\eta-{1\over2}m^2\eta^2\bigg)~,
\eeq
where $m$ denotes the mass of the scalar field, which includes its bare mass as well as quantum corrections, and all other quantities are those defined earlier in this work. When compactifying 3 of the dimensions parallel to the brane as before, the scalar field will be decomposed in its Kaluza-Klein (KK) modes, leading to a tower of 4-dimensional massive modes, or KK states. We will focus on the evolution of the zero mode of the field $\eta$, whose mass will simply be given by $m$. To simplify the notation, we will denote this mode as $\eta$, although one must bear in mind that it is not the original 7-dimensional field. The global factor of $T_6$ will not affect our discussion and we may absorb it into the definition of the field.

First, let us consider the case $m=0$. Then, the classical equations of motion arising from the action Eq. (\ref{eq43}) are given by:
\beq \label{eq44}
\partial_{\mu}(e^{-\Phi}\sqrt{\hat\gamma}\hat\gamma^{\mu\nu}\partial_{\nu}\eta)=0~.
\eeq

Using the results obtained in Section II, and defining the function $f(t)\equiv[h(1-hv^2)]^{1\over2}$, we can write this as
\beq \label{eq45}
\ddot\eta+(3H-F)\dot\eta-{1\over a^2}(1-hv^2)\nabla^2\eta=0~,
\eeq
where $F={\dot f\over f}$ and $\nabla^2\eta\equiv\eta^{ij}\partial_i\partial_j\eta$ is the flat 3-dimensional Laplacian of the field.

To construct the associated quantum description, we follow the semi-classical approach to the quantization of scalar fields in curved space \cite{Birrell}. In a curved background spacetime with a time-varying geometry, as is the case of the world-volume of the probe brane, the induced time-variation of the fields modifies the usual canonical quantization procedure, as the quantum operators become themselves time-dependent. In particular, the creation and annihilation operators associated with the field will now evolve in time as the background changes, the same happening with the associated multi-particle states. This is the main reason behind the production of particles in a dynamical background. Let us start by expanding the field in Fourier modes of the form:
\beq \label{eq46}
\eta(\mathbf{x},t)=\int {d^3k\over(2\pi)^{3\over2}}\big(a_{\mathbf{k}}\chi_{\mathbf{k}}(t)e^{i\mathbf{k}\cdot\mathbf{x}}+ a^{\dagger}_{\mathbf{k}}\chi^*_{\mathbf{k}}(t)e^{-i\mathbf{k}\cdot\mathbf{x}}\big)~.
\eeq

In the quantum theory, $a_{\mathbf{k}}$ and $a^{\dagger}_{\mathbf{k}}$ become the annihilation and creation operators associated with the Fourier mode $\mathbf{k}$ of the field. Expanding the field in this way, we include all the time dependence of the field in the functions $\chi_{\mathbf{k}}(t)$ and $\chi^*_{\mathbf{k}}(t)$, while the operators remain time-independent. The conjugate momentum to the field $\eta$, obtained by computing $\delta S_{\eta}/\delta \dot\eta$, can then be written as
\beq \label{eq47}
\pi(\mathbf{x},t)={a^3\over f}\int{d^3k\over(2\pi)^{3\over2}}\big(a_{\mathbf{k}}\dot\chi_{\mathbf{k}}e^{i\mathbf{k}\cdot\mathbf{x}}+ a^{\dagger}_{\mathbf{k}}\dot\chi^*_{\mathbf{k}}e^{-i\mathbf{k}\cdot\mathbf{x}}\big).
\eeq

If the creation and annihilation operators satisfy the canonical commutation relations
\beqa \label{eq48}
[a_{\mathbf{k}},a_{\mathbf{k'}}]=[a^{\dagger}_{\mathbf{k}},a^{\dagger}_{\mathbf{k'}}]=0~,\ \ [a_{\mathbf{k}},a^{\dagger}_{\mathbf{k'}}]=\delta^3(\mathbf{k}-\mathbf{k'})~,
\eeqa
then the canonical commutators for the field and its conjugate momentum, $[\eta(\mathbf{x},t),\pi(\mathbf{y},t)]=i\delta^3(\mathbf{x}-\mathbf{y})$, can only be obtained if the mode functions satisfy the following Wronskian normalization condition:
\beq \label{eq49}
\chi_{\mathbf{k}}\dot\chi^*_{\mathbf{k}}-\chi^*_{\mathbf{k}}\dot\chi_{\mathbf{k}}=i{f\over a^3}~.
\eeq

This condition is essential for consistency of the quantization procedure. We now follow a semi-classical approach and consider the evolution of the field modes to be given by the classical equations of motion, assuming that any quantum corrections to their propagation can be neglected. Substituting Eq. (\ref{eq46}) into Eq. (\ref{eq45}), we see that each of the Fourier modes evolves independently according to
\beq \label{eq50}
\ddot\chi_{\mathbf{k}}+(3H-F)\dot\chi_{\mathbf{k}}+{k^2\over a^2}(1-hv^2)\chi_{\mathbf{k}}=0~.
\eeq
and similarly for the complex conjugate modes $\chi^*_{\mathbf{k}}$. Notice that this equation depends only on $k^2\equiv|\mathbf k|^2$ but not the direction of the momenta $\hat{k}\equiv{\mathbf{k}\over k}$. This is a consequence of the isotropy of our metric \textit{ansatz}, and we may write without loss of generality $\chi_{\mathbf{k}}(t)=\chi_k(t)$. It is also clear, from Eq. (\ref{eq50}), that the physical momentum of the modes is related to their comoving momentum $k$ via
\beq \label{eq51}
k_{phys}={k\over a}\sqrt{1-hv^2}~.
\eeq
This expression includes both the expected redshift of the modes in an expanding universe and the effect of the probe's motion through the dynamical background.

Let us now rescale the field modes by defining the mode functions $X_k(t)\equiv a^{3\over2}(t)f^{-{1\over2}}(t)\chi_k(t)$. These satisfy
\beq \label{eq52}
\ddot{X}_k+\omega_k^2X_k=0~,
\eeq
which corresponds to the equation for a harmonic oscillator with a variable frequency given by:
\beqa \label{eq53}
\omega_k^2&=&{k^2\over a^2}(1-hv^2)-{1\over4}(3H-F)^2-{1\over2}(3\dot H-\dot F)=\nonumber\\
&=&k_{phys}^2+\Delta^2~,
\eeqa
where we have defined
\beq \label{eq54}
\Delta^2(t)\equiv-{1\over4}(3H-F)^2-{1\over2}(3\dot H-\dot F)~.
\eeq

As each mode behaves as a harmonic oscillator, we can define the associated particle number via
\beqa \label{eq55}
E_k&=&{1\over2}|\dot X_k|^2+{1\over2}\omega_k^2|X_k|^2=\omega_k\bigg(n_k+{1\over2}\bigg)\nonumber\\
\Leftrightarrow n_k&=&{\omega_k\over2}\bigg({|\dot X_k|^2\over\omega_k^2}+|X_k|^2\bigg)-{1\over2}~.
\eeqa

It is well known that this quantity is an adiabatic invariant for a harmonic oscillator with variable frequency, as happens, for example, in an oscillating pendulum whose length is decreased infinitely slowly \cite{Birrell}. Thus, the number of \textit{quanta} in a given Fourier mode $k$ can only change if its frequency varies in a non-adiabatic way, which can be expressed by the following condition \cite{Kofman}:
\beq \label{eq56}
\left|{d\omega_k\over dt}\right|\gsim \omega_k^2~.
\eeq

For our particular case, this gives, in the limit $hv^2\ll1$,
\beq \label{eq57}
|Hk_{phys}^2-\Delta\dot\Delta|\gsim (k_{phys}^2+\Delta^2)^{3\over2}~.
\eeq
This condition will then constrain the number of particles produced in each physical momentum mode by the motion of the probe brane.

To quantify this number of particles, we need to look for solutions of the mode equation, Eq. (\ref{eq52}). We will start by discussing what happens in a non-expanding universe and then we will analyze the effects of Hubble expansion on particle production.


\subsection{Non-expanding universe}

Consider the motion of the probe brane in a non-expanding universe, $H=0$, and assume that the non-relativistic and large distance approximations discussed earlier are valid. As we have seen in Section II, the probe follows closed elliptical trajectories if its energy (density) is negative, $\epsilon<0$. If $hv^2\ll1$, we have to leading order:
\beq \label{eq58}
1-hv^2\simeq1~,\qquad k_{phys}\simeq k~, \qquad f\simeq h^{1/2}~.
\eeq
Then, we may take the following approximations:
\beqa \label{eq59}
F&=&{\dot f\over f}\simeq{1\over2}{\dot h\over h}~,\nonumber\\
\dot F&\simeq&{1\over2}\bigg[{\ddot h\over h}-\bigg({\dot h\over h}\bigg)^2\bigg]~,
\eeqa
so that we can write:
\beq \label{eq60}
\Delta^2=-{1\over4}F^2+{1\over2}\dot F\simeq-{5\over16}\bigg({\dot h\over h}\bigg)^2+{1\over4}{\ddot h\over h}~.
\eeq

An exact analytical expression for the time variation of the radial field $\rho(t)$ can be obtained in a non-expanding universe, being given implicitly by
\beq \label{eq61}
\rho(t)=R\big[1-e\cos\big(\psi(t)\big)\big]~.
\eeq
The angular variable $\psi(t)$ satisfies
\beq \label{eq62}
t-t_0=\Omega^{-1}\big[\psi-\psi_0+e(\sin\psi-\sin\psi_0)\big]~,
\eeq
where $\psi_0\equiv \psi(t_0)$ and $\Omega=\sqrt{\sigma\over2}R^{-{3\over2}}$ is the angular frequency of the orbit. The angular variables $\psi$ and $\theta$ are related by:
\beq \label{eq63}
\tan{\theta\over2}=\sqrt{1+e\over1-e}\tan{\psi\over2}~.
\eeq

For small eccentricity orbits, $e\ll1$, Eq. (\ref{eq62}) gives, to leading order in $e$:
\beq \label{eq64}
\psi\simeq\psi_0+\Omega(t-t_0)~.
\eeq
Choosing $\psi_0=\pi$ we have, in this approximation,
\beq \label{eq65}
\rho(t)\simeq R\big[1+e\cos\big(\Omega(t-t_0)\big)\big]~.
\eeq

This allows us to determine the non-relativistic and large distance expression for $\Delta^2$, in the case of small eccentricity orbits. To lowest order, we then obtain:
\beq \label{eq66}
\Delta^2\simeq{1\over4}\sqrt{T_6V_3\over2}{Q_6\over R}e\Omega^2\cos\big(\Omega(t-t_0)\big)~.
\eeq 
Defining 
\beq \label{eq67}
\delta^2\equiv{1\over4}\sqrt{T_6V_3\over2}{Q_6\over R}e\Omega^2~,
\eeq
the variable frequency of the mode with comoving momentum $k$ can, hence, be approximated by:
\beq \label{eq68}
\omega_k^2(t)\simeq k^2+\delta^2\cos\big(\Omega(t-t_0)\big)~.
\eeq

If we now rewrite Eq. (\ref{eq52}) in terms of the rescaled time variable $z\equiv{\Omega\over2}(t-t_0)$, we obtain:
\beq \label{eq69}
X^{\prime\prime}_k+\big(A_k-2q\cos(2z)\big)X_k=0~,
\eeq
where
\beq \label{eq70}
A_k\equiv{4k^2\over\Omega^2}~,\qquad q\equiv-2{\delta^2\over\Omega^2}~.
\eeq

Eq. (\ref{eq69}) has the form of the well-known Mathieu equation \cite{McLachlan} with parameters $A_k$ and $q$ determined by the orbital parameters of the probe brane and by the value of the comoving momentum $k$ of each mode. In the parameter space $(A_k,q)$, the Mathieu equation exhibits both stable and unstable solutions, the latter being closely associated with the phenomenon of parametric resonance \cite{Landau}. As we are working under the assumption that the probe brane's orbit has a large radius and a small eccentricity, it is easy to check that $|q|\ll1$ (the analysis is independent of the sign of $q$). In this region of parameter space, the Mathieu equation exhibits instabilities in a series of narrow resonance bands near $A_k\sim n$, with $n$ being a positive integer, and with a width in comoving momentum space approximately given by $\Delta k^{n}\sim|q|^n$. In these resonance bands, the solution evolves as $X_k\propto e^{\mu_k^{(n)}z}$, with a real exponent $\mu_k^{(n)}$, according to Floquet's Theorem.

The most important of these is the first resonance band, which occurs when 
\beq \label{eq71}
1-|q|-{1\over8}q^2\lsim A_k\lsim 1+|q|-{1\over8}q^2~.
\eeq 

The exponent $\mu_k\equiv\mu_k^{(1)}$ can then be approximately written as
\beq \label{eq72}
\mu_k\simeq{1\over2}\sqrt{q^2-(A_k-1)^2}~.
\eeq

This is, as claimed, real for $1-|q|\lsim A_k \lsim 1+|q|$, having a maximum value of $|q|/2$ for $A_k\simeq 1$. This implies that particle-modes lying inside this resonance band will be exponentially amplified, leading to a resonant production of particles. Thus, from Eq. (\ref{eq70}), the center of the resonance band occurs for comoving momentum $k_c\simeq\Omega/2$, while its lower and upper bounds are given by
\beqa \label{eq73}
k_{min}&\simeq&\sqrt{\bigg({\Omega\over2}\bigg)^2-{\delta^2\over2}}\simeq k_c\bigg[1-\bigg({\delta^2\over\Omega}\bigg)^2\bigg]~,\nonumber\\
k_{max}&\simeq&\sqrt{\bigg({\Omega\over2}\bigg)^2+{\delta^2\over2}}\simeq k_c\bigg[1+\bigg({\delta^2\over\Omega}\bigg)^2\bigg]~,
\eeqa
so that the resonance band has a width in comoving momentum space of $\Delta k_{res}\equiv k_{max}-k_{min}\simeq{\delta^2\over\Omega}$. This is quite small compared to the value of the center of the resonance, as ${\Delta k_{res}\over k_c}\simeq|q|\ll1$, which means that particle production occurs in a regime of narrow parametric resonance. The particles produced by this mechanism have typical energies of order $\omega_k\simeq{\Omega\over2}$, i.e. with half the typical energy of the field $\phi$. This agrees with the discussion in \cite{Burgess}, where it is stated that particles are produced in pairs with opposite momenta, as if resulting directly from the decay of the interbrane distance field\footnote{This field can be seen as a classical condensate of zero momentum particles, so that only pair production of $\eta$-particles ensures momentum conservation.}. We have found in this work, however, that these particles are produced resonantly, the associated field modes being exponentially amplified.

After the modes inside the resonance band have been amplified sufficiently, we may write, from Eq. (\ref{eq55}),
\beq \label{eq74}
n_k\propto e^{2\mu_kz}=e^{\mu_k\Omega(t-t_0)}~,
\eeq
where we used that $\mu_k\lsim{|q|\over2}\ll1$, so that the term involving $|\dot X_k|^2$ in Eq. (\ref{eq55}) can be neglected. This also implies that, although the particle number in each mode grows exponentially, the resonance takes a long time to develop. In particular, the typical resonance time, $\Delta t_{res}=(\mu_k\Omega)^{-1}$, is much larger than the orbital period of the probe brane, which is of order $2\pi\Omega^{-1}$. This is characteristic of the narrow resonance regime and, hence, a significant particle number is only produced after the probe brane has completed a large number of orbits, given by $(2\pi\mu_k)^{-1}$. 

To have an idea of the order of magnitude of these quantities, consider the scenario studied in Section II, where $V_3=T_6^{-1}=(2\pi)^6$ with $Q_6=100$. Then, for an orbit with eccentricity $e=0.01\ll1$ and an angular momentum $l=500$, we have $\sqrt{T_6V_3/2}{Q_6\over R}\simeq 0.03$, which satisfies the large distances approximation, and $|q|\simeq 10^{-4}\ll1$. This implies that the probe brane will have to complete approximately $3\times10^3$ orbits for the particle number to be amplified by one e-folding.

To illustrate the results obtained earlier, we have solved Eq. (\ref{eq52}) for the example described above. An initial vacuum state with zero particle number corresponds to the following initial conditions:
\beq \label{eq75}
X_k(t_0)={1\over\sqrt{2\omega_k(t_0)}}~,\qquad \dot{X}_k(t_0)=-i\sqrt{{\omega_k(t_0)\over2}}~.
\eeq

In Figure 2 we have plotted the numerical results obtained for the comoving momentum at the center of the resonance band, $k_c$. The linear evolution observed for the logarithm of the particle number shows that, as expected, this quantity is being amplified exponentially with time, but that a significant number of particles is produced only after the probe brane has completed a few thousands of orbits.

\begin{figure}[htbp]
	\centering
		\includegraphics[scale=0.8]{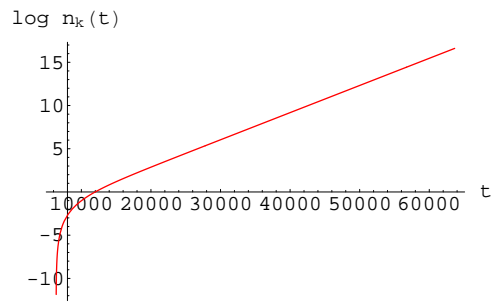}
	\label{fig:figure_2}
	\caption{Numerical results obtained for the evolution of the particle number associated with the mode in the center of the resonance band for an orbit with $e=0.01$ and $l=500$, giving $q\simeq10^{-4}$. The time coordinate $t$ is given in units of the orbital period $T=2\pi\Omega^{-1}\simeq1.57\times10^5$.}
\end{figure}

We note that, although this solution was obtained numerically, a formal analytical solution of the Mathieu equation can be written in terms of the Mathieu sine and cosine functions. 

In Figure 3 we also show the evolution of the particle number for a mode outside the limits of the resonance band. In this case, $\mu_k$ takes imaginary values and $n_k$ exhibits an oscillating behavior with a small amplitude, confirming that only for comoving momenta $k_{min}\leq k\leq k_{max}$ a significant particle number is produced by the elliptical motion of the probe brane. It is also worth mentioning that, in the limit of circular orbits, $e\rightarrow0$, the parameter $q$ vanishes and there is no amplification of any of the particle momentum modes. This confirms our early assumption that the interbrane brane distance needs to vary for particle production to occur.

\begin{figure}[htbp]
	\centering
		\includegraphics[scale=0.8]{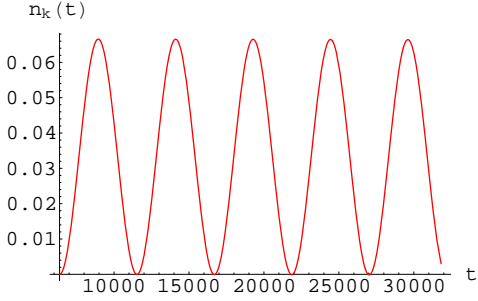}
	\label{fig:figure_3}
	\caption{Evolution of the particle number for a particle-mode outside the resonance band for an orbit with $e=0.01$ and $l=500$, giving $q\simeq10^{-4}$. The comoving momentum of the mode is $k=k_c\big(1+4{\delta^2\over\Omega^2}\big)>k_{max}$. The time coordinate $t$ is given in units of the orbital period $T=2\pi\Omega^{-1}\simeq1.57\times10^5$.}
\end{figure}

So far we have not included the backreaction effects of the produced particles on the probe's motion. We will return to this later in section III C, where we estimate the energy density damped into brane-modes by the resonance and discuss how this affects the probe brane's orbital decay. 

Consider now the case where the field has a non-vanishing mass, $m\neq0$. In this case, the classical equations of motion Eq. (\ref{eq44}) can be generalized to
\beq \label{eq76}
\partial_{\mu}(e^{-\Phi}\sqrt{\hat\gamma}\hat\gamma^{\mu\nu}\partial_{\nu}\eta)-e^{-\Phi}\sqrt{-\hat\gamma}m^2\eta=0~,
\eeq
or, explicitly,
\beq \label{eq77}
\ddot\eta+(3H-F)\dot\eta-{1\over a^2}(1-hv^2)\nabla^2\eta+{(1-hv^2)\over h^{1\over2}}m^2\eta=0~.
\eeq

Each of the Fourier modes then evolves independently according to
\beq \label{eq78}
\ddot\chi_k+(3H-F)\dot\chi_k+\bigg[(1-hv^2){k^2\over a^2}+{(1-hv^2)\over h^{1\over2}}m^2\bigg]~.
\eeq

This implies that all modes have, in this case, an effective physical mass which is different from the mass parameter $m$ due to the motion of the probe brane:
\beq \label{eq79}
m_{eff}^2={(1-hv^2)\over h^{1\over2}}m^2~.
\eeq
Notice that the mass is not, however, affected by the expansion of the universe. When writing the equations of motion in terms of the rescaled mode function $X_k(t)$ defined earlier, we again obtain the equation for a harmonic oscillator with a varying frequency, which is now given by:
\beq \label{eq80}
\omega_k^2=k_{phys}^2+m_{eff}^2+\Delta^2~,
\eeq 
where all the quantities are time-dependent in the general case. Let us now set $H=0$ and compare these results with the ones obtained for the particle production mechanism in the massless case. First, notice that, in the non-relativistic and large distance approximation, we have $m_{eff}^2\simeq m^2$. Thus, the leading order modification introduced by the mass of the field can be obtained by simply replacing $k^2\rightarrow k^2+m^2$. 

Following the same procedure as before, one can reduce the particle-mode equation to the Mathieu equation, with the parameter $A_k$ being now given by:
\beq \label{eq81}
A_k={4(k^2+m^2)\over\Omega^2}~.
\eeq

The parameter $q$, which quantifies the strength and width of the resonance phenomenon, is not modified by the introduction of the mass parameter, so that we expect the particle production mechanism to occur as before. In this case, however, the center of the resonance band is shifted to a lower momentum, $k_c=\sqrt{({\Omega/2})^2-m^2}$, which gives $A_k=1$. The upper and lower limits of the resonance band are also modified, being now given by:
\beqa \label{eq82}
k_{min}=\sqrt{({\Omega/2})^2-m^2-{1\over2}\delta^2}~,\nonumber\\
k_{max}=\sqrt{({\Omega/2})^2-m^2+{1\over2}\delta^2}~,
\eeqa
which in turn alters the width of the resonance band in the obvious way. If, as in the massless case, we have $\delta^2/2\ll k_c^2$, we may write $\Delta k_{res}\simeq {\delta^2/2k_c}$, which is larger than the corresponding value in the massless case, satisfying nevertheless the narrow resonance condition. Notice that this implies an upper bound for the mass of the particles which can be produced by this mechanism, 
\beq \label{eq83}
m\leq\sqrt{\bigg({\Omega\over2}\bigg)^2-{1\over2}\delta^2}\simeq{\Omega\over2}~.
\eeq
In the particle description discussed earlier, this simply means that each ``particle" in the interbrane distance field $\phi$ needs to have enough energy to decay into two $\eta$-particles at rest. A residual kinetic energy is also necessary if all the modes inside the resonance band are to be excited. In the non-relativistic and large distance approximation, the orbital frequency of the probe brane is assumed to be small, so that only particles with small masses may be excited during its motion. One must recall, however, that we are measuring all quantities in terms of the typical string parameters. In particular, the orbital frequency should be small compared to the string energy scale. If the latter corresponds to the Planck scale, then the produced particles can still be quite massive.

For the example studied earlier in this section, with orbital parameters $\Omega\simeq 4\times10^{-5}$, particles up to masses of order $10^{13}$ GeV can be resonantly created. Recalling the definition of $\phi$ in terms of the physical interbrane distance, Eq. (\ref{eq17}), we may write the upper bound on the mass of the produced particles in terms of the physical semi-major axis of the orbit, $R_{phys}=\sqrt{2/T_6V_3}R$. Inserting back the missing $l_s$ factors, we obtain:
\beq \label{eq84}
m_{max}=\sqrt{Q_6\over2R_{phys}}{l_s\over R_{phys}}M_s~,
\eeq
where $M_s=l_s^{-1}$ is the string energy scale. As we assumed in the beginning that $r\gg Q_6\gg l_s$, it is clear from this expression that $m_{max}\ll M_s$, which allows nevertheless for the production of very massive particles. In general, particles living in the world-volume of the probe brane will be massive, so that we need to use the results obtained above to describe the parametric resonance. For masses which are small compared to the typical orbital energies of the probe, the simpler treatment of the $m=0$ case will be sufficient.


\subsection{Expanding universe}

Let us now consider the case of an expanding universe, with $H={\alpha\over t}>0$. The inclusion of the scale factor will modify our previous analysis of particle production in several ways. First, it will change the frequency of each harmonic oscillator particle-mode, namely modifying the factor $\Delta^2(t)$, given in Eq. (\ref{eq54}). Also, the physical momenta of the particle-modes will be redshifted as the universe expands, so that, in the non-relativistic and large distance approximation, we have $k_{phys}\simeq k/a\propto (t/t_0)^{-\alpha}$. Finally, the motion of the probe brane will be altered, as we have seen in Section II, so that the semi-major axis of elliptical orbits will decrease as $R(t)\propto({t/t_0})^{-6\alpha}$, making the angular frequency consequently increase as $\Omega(t)\propto({t/t_0})^{9\alpha}$. Furthermore, if the Hubble parameter is sufficiently large, the orbital eccentricity may exhibit a significant growth.


\subsubsection{Analytical results}

All the modifications make a complete analytical study of particle production in the probe brane rather complex. In order to determine the leading effects of the expansion, we will consider the motion of the probe brane at sufficiently late times, so that some of the effects of the expansion can be discarded. As we have concluded in Section II, the variation of the eccentricity of the orbits is, in this regime, very small, so that it may be neglected. The explicit effects of the expansion in modifying the factor $\Delta^2$ may also be discarded in this limit, provided that $|F|\gg 3H$ and $|\dot F|\gg|3\dot H|$, or explicitly
\beq \label{eq85}
t\gg {6\alpha\over e\Omega}\bigg({R_{phys}\over Q_6}\bigg)~,\qquad t^2\gg {6\alpha\over e\Omega^2}\bigg({R_{phys}\over Q_6}\bigg)~.
\eeq
The first of these conditions will, \textit{a priori}, be more constraining. For the example we have been following in this section, the first condition gives $t\gg 5\times 10^8$ while the second only implies the constraint $t\gg 3.5\times 10^6$.

With these constraints in mind, we now have to analyze how Hubble expansion modifies the factors $F$ and $\dot F$ in the variable harmonic oscillator frequency. The radial interbrane distance field $\rho$ can now be written, in the low eccentricity orbit approximation, as:
\beq \label{eq86}
\rho(t)\simeq R(t)\big[1+e\cos\big(\Omega(t)(t-t_0)\big)\big]~,
\eeq
with
\beq \label{eq87}
R(t)=R_0\bigg({t\over t_0}\bigg)^{-6\alpha}~, \qquad \Omega(t)=\Omega_0\bigg({t\over t_0}\bigg)^{9\alpha}~.
\eeq

Computing its time derivatives, we find that they involve terms arising from Hubble expansion which are suppressed by at least one power of $t$, being negligible at late times. There are also terms involving the quantity ${\Delta t\over t_0}={t-t_0\over t_0}$. If the time necessary for particle production to occur is smaller than the age of the universe at that time, we may also discard these terms. After some algebra, the leading order modification to $\Delta^2$ is then given by
\beq \label{eq88}
\Delta^2(t)\simeq \delta^2(t)\cos\big(\Omega(t)(t-t_0)\big)~.
\eeq
The quantity $\delta^2$ is now time-dependent and can be obtained trivially by including the appropriate time variation of the orbital parameters in its non-expanding expression, Eq. (\ref{eq67}). This gives
\beq \label{eq89}
\delta^2(t)=\delta_0^2\bigg({t\over t_0}\bigg)^{24\alpha}~,\qquad \delta_0^2\equiv {1\over4}\sqrt{T_6V_3\over2}{Q_6\over R_0}e\Omega_0^2~.
\eeq

Then, taking into account the momentum redshift, we can write the particle-mode equation, to lowest order, as:
\beq \label{eq90}
\ddot{X}_k+\bigg[k^2\bigg({t\over t_0}\bigg)^{-2\alpha}+\delta_0^2\bigg({t\over t_0}\bigg)^{24\alpha}\cos\big(\Omega(t)(t-t_0)\big)\bigg]X_k=0~.
\eeq

Following the same reasoning as in the non-expanding case, we rescale the time coordinate by defining the variable $z\equiv{\Omega(t)\over2}(t-t_0)$. As the angular frequency of the orbit is now time-dependent, this change of variables will introduce new terms in the equation. These will, however, be suppressed by powers of $t$ and involve first and second derivatives of the mode function $X_k$. In the non-expanding case, these were quite small compared to $X_k$, and one expects them to be negligible in this case as well. Using that, to lowest order, $z\simeq{\Omega_0\over2}\Delta t$, we may write the leading order equation for the mode function in terms of the variable $z$ as
\beq \label{eq91}
X^{\prime\prime}_k+[A_k(z)-2q(z)\cos(2z)]X_k=0~,
\eeq
with
\beq \label{eq92}
A_k(z)\equiv A_{k0}(1-\gamma z)~, \qquad q(z)\equiv q_0(1+\xi z)~,
\eeq
where $A_{k0}\equiv4k^2/\Omega_0^2$ and $q_0\equiv-2\delta_o^2/\Omega_0^2$ correspond to the initial values of the parameters of the Mathieu equation, given by their non-expanding case expressions, and the coefficients $\gamma$ and $\xi$ are given by
\beq \label{eq93}
\gamma\equiv {40\alpha\over\Omega_0t_0}~, \qquad \xi\equiv{12\alpha\over\Omega_0t_0}={3\over10}\gamma~.
\eeq

We, hence, conclude that the leading order effect of the expansion is to make the coefficients of the Mathieu equation vary in time. In particular, their variation is, to a first approximation, linear in the variable $z$ and controlled by the single parameter $\gamma$, which will be small at late times. One must take into account, however, that Eq. (\ref{eq91}) is not the Mathieu equation and, namely, does not satisfy the conditions of Floquet's Theorem which allowed us to define the coefficient $\mu_k$. Nevertheless, we can use the results obtained in the non-expanding case from the properties of the Mathieu equation as a guide to describe the behavior of the modes when $H>0$.

First, notice that, as $A_k$ decreases with $z$, the position of the center of the resonance band, at $A_k=1$, will be shifted to higher momentum modes:
\beq \label{eq94}
k_c(z)={k_{c0}\over1-\gamma z}~, \qquad k_{c0}\equiv{\Omega_0\over2}~. 
\eeq
Thus, we expect modes with comoving momentum larger than $k_{c0}$ to be excited as the probe's orbit decays. To better understand the evolution of each mode, let us analyze the evolution of the exponent $\mu_k(z)$, which is given by:
\beq \label{eq95}
\mu_k(z)\simeq{1\over2}\sqrt{q_0^2(1+\xi z)^2-(A_{k0}(1-\gamma z)-1)^2}~.
\eeq
The zeros of $\mu_k(z)$ occur for
\beq \label{eq96}
z_1={-|q_0|+A_{k0}-1\over\gamma A_{k0}+\xi |q_0|}~,\qquad z_2={|q_0|+A_{k0}-1\over\gamma A_{k0}-\xi |q_0|}~,
\eeq
with $z_2>z_1$. This means that $\mu_k(z)$ will be real for $z_1\leq z\leq z_2$ and pure imaginary for $z<z_1$ and $z>z_2$. Figure 4 illustrates the typical evolution of $\mu_k(z)$ for a generic Fourier mode. 

\begin{figure}[htbp]
	\centering
		\includegraphics[scale=0.7]{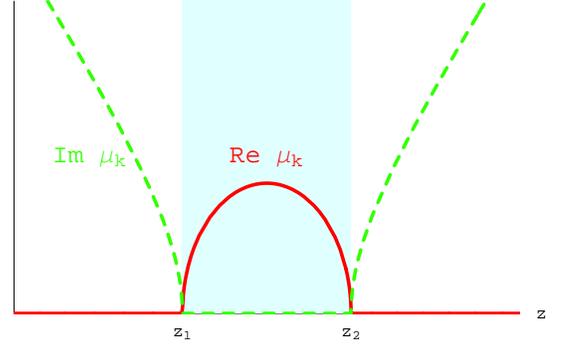}
	\label{fig:figure_4}
	\caption{Evolution of the real and imaginary parts of the exponent $\mu_k(z)$ for a generic mode. The shaded area corresponds to the period the mode spends inside the resonance band.}
\end{figure}

Thus, each mode may only experience the resonant regime during a finite amount of time, when $\mu_k(z)$ is real and the mode is inside the resonance band. During the periods where $\mu_k(z)$ is pure imaginary, we expect the particle number to exhibit an oscillating behavior. In the non-expanding case, the frequency and amplitude of these oscillations for modes outside the resonance band is determined by the value of $\mathrm{Im}\ \mu_k$, so that the larger the latter the smaller the period and the amplitude of the oscillations. As, in an expanding universe, $\mathrm{Im}\ \mu_k(z)$ changes in time, we expect the frequency and amplitude of the oscillations of the particle number to vary as the probe brane's orbit decays. Similarly, as $\mathrm{Re}\ \mu_k$ is not constant when the mode is inside the resonance band, the strength of the resonance is also expected to vary.

Particle modes will then have different behaviours according to their position relative to the initial resonance band, which can be parametrized by:
\beq \label{eq97}
A_{k0}\equiv1+\beta |q_0|~.
\eeq

Recalling that the centre of the original resonance band corresponds to $\beta=0$ and that the probe's motion begins at $z=0$, we need to distinguish three different types of modes:

\begin{enumerate}
\item[\textbf{I.}] $\beta<-1$

As both $z_1$ and $z_2$ are negative in this case, these modes will always be outside the resonance band, with an imaginary exponent $\mu_k(z)$ for all $z\geq0$. 

\item[\textbf{II.}] $-1\leq \beta\leq 1$

This case corresponds to the particle-modes inside the resonance band in a non-expanding universe, with $z_1\leq0$ but $z_2\geq0$. Hence, they will start inside the resonance band but stop experiencing the resonant regime for $z>z_2$.

\item[\textbf{III.}] $\beta>1$

Although these modes are outside the initial resonance band, they will experience the resonant behavior at some later stage, as $z_2>z_1>0$.
\end{enumerate}

For both type II and type III modes, the maximum value of $\mathrm{Re}\ \mu_k(z)$ occurs for $z_{max}={z_1+z_2\over2}$, giving, for $|q_0|\ll1$, 
\beq \label{eq98}
\mu_k(z_{max})\simeq {|q_0|\over2}\bigg(1+{3\over10}\beta|q_0|\bigg)~.
\eeq

This gives the same result as in the non-expanding limit for $\beta=0$. It is easy to check that $z_{max}\rightarrow\pm\infty$ as $\gamma\rightarrow0$, the same happening for $z_1$ and $z_2$. This implies that only modes for which $z_1\leq0$ ($\rightarrow -\infty$) and $z_2\geq0$ ($\rightarrow+\infty$) will be excited in this limit, in agreement with the fact that only type II modes are excited in the non-expanding case ($\gamma=0$) and that these experience the parametric resonance during an infinite period of time. It is also clear from Eq. (\ref{eq98}) that higher momentum modes are excited by the resonance at later times, as expected from the evolution of the center of the resonance band.

The total time each type II or type III mode spends inside the resonance band is given by:
\beq \label{eq99}
\Delta z_{band}\equiv z_2-z_1\simeq{2|q_0|\over\gamma}\bigg(1-{7\over10}\beta|q_0|\bigg)~.
\eeq 

It is clear that this quantity decreases with $\gamma$, so that the smaller the Hubble parameter the more time each mode is excited by the resonance. In the limit $\gamma\rightarrow0$, $\Delta z_{band}\rightarrow+\infty$, but as previously discussed this limit only applies to type II modes, as all the others are outside the resonance band in this limit. One also concludes that higher momentum modes will spend less time inside the resonance band, which suggests these will be less excited.

The analysis of the properties of the exponent $\mu_k(z)$ thus provides a very clear insight on the qualitative evolution of all momentum modes. We would like, however, to be able to quantify the particle number produced by the resonance in each mode. In the non-expanding case, $X_k(z)\propto e^{\mu_k z}$, which implied $n_k(z)\propto e^{2\mu_kz}$ in the limit of large particle number for type II modes. Despite the variation of $\mu_k$ in the $H>0$ case, we may take this exponent to be approximately constant during an infinitesimal interval between $z$ and $z+dz$. Then, during this interval, $X_k$ will be amplified by a factor $\exp(\mu_k(z)dz)$ if the mode is inside the resonance band. Integrating this result, we expect that, after a significant number of particles has been produced,
\beq \label{eq100}
n_k(z)\propto \exp\bigg(2\int_{z_i}^{z}\mu_k(z')dz'\bigg)~.
\eeq
The initial time $z_i$ refers to the time the mode enters the resonance band, i.e. $z_i=0$ and $z_i=z_1$ for type II and type III modes, respectively. One can use Eq. (\ref{eq100}) to estimate the total particle number produced in each mode by the parametric resonance. The details of this calculation are given in Appendix A. For type II and type III modes, one obtains:
\beqa \label{eq101}
\log n_k^{res (II)}(z_2)&=&{q_0^2\over2\gamma}{1\over1+\beta |q_0|}\big(\beta\sqrt{1-\beta^2}+{\pi\over2}+\nonumber\\
&+&\arcsin\beta\big)~,\nonumber\\
\log n_k^{res (III)}(z_2)&=&{q_0^2\over2\gamma}{\pi\over1+\beta |q_0|}~.
\eeqa

Eq. (\ref{eq101}) constitutes the main result of this section, giving the leading order expressions for the particle number density produced by the resonance in an expanding universe. From these approximated expressions, we can see that the parameter $q_0^2/\gamma$ controls the strength of the resonance, a significant number of particles being produced only if $\gamma\lsim q_0^2$. Recall that, in the non-expanding case, the typical resonance time was given by $\Delta t_{res}=(\mu_k\Omega)^{-1}\sim(|q_0|\Omega_0/2)^{-1}$. The typical time a mode spends inside the resonance band in the expanding case is, from Eq. (\ref{eq99}), of order $\Delta t_{band}\sim {4q_0\over\gamma}\Omega_0^{-1}$. Hence, apart from numerical factors, the condition for a significant number of particles to be produced is simply stating that the time a given mode spends inside the resonance band should be greater than the typical time the resonance takes to develop, $\Delta t_{band}\gsim \Delta t_{res}$. 

From the results in Eq. (\ref{eq101}), we can also conclude that the total particle number produced increases with $\beta$ for $-1\leq\beta\leq1$. On the other hand, for $\beta>1$, it exhibits a slow decrease with $\beta$, showing that the production of high momentum modes is suppressed. This agrees with the fact that high momentum modes spend less time inside the resonance band, as obtained above.


\subsubsection{Numerical simulations in an expanding universe}

To check the results obtained so far for the resonant particle production in an expanding universe, we have solved Eq. (\ref{eq91}) numerically. In order to simplify the computation, we have rescaled the mode functions via $\tilde{X}_k\equiv\sqrt{2\omega_k(t_0)}X_k$, so that initially we have approximately
\beq \label{eq102}
\tilde{X}_k(0)\simeq1~,\qquad \tilde{X}^{\prime}_k(0)\simeq-i~.
\eeq
We then computed the function
\beq \label{eq103}
\tilde{n}_k(z)=|\tilde{X}_k(z)|^2+|\tilde{X}^{\prime}_k(z)|^2~.
\eeq
This is related to the physical particle number density of each mode, given by Eq. (\ref{eq55}), approximately by $n_k\simeq{\tilde{n}_k\over4a(t)}$, taking the limit of large particle number. Note that $\tilde{n}_k=2$ initially, as we neglected the $1/2$ factor in Eq. (\ref{eq55}), so that the effects of this normalization have to be taken into account before the resonance produces a significant number of particles.

In Figure 5, we illustrate the results obtained for modes of types I, II and III, including also the analytical prediction obtained from Eq. (\ref{eq100}). These simulations correspond to an original resonance strength $|q_0|=10^{-3}$, the effects of the expansion being quantified by $\gamma=10^{-7}$ (recall Eq. (\ref{eq93})). These values ensure the validity of the approximations made and also that a significant particle number is produced for modes of types II and III.

\begin{figure}[htbp]
	\centering
		\includegraphics[scale=0.8]{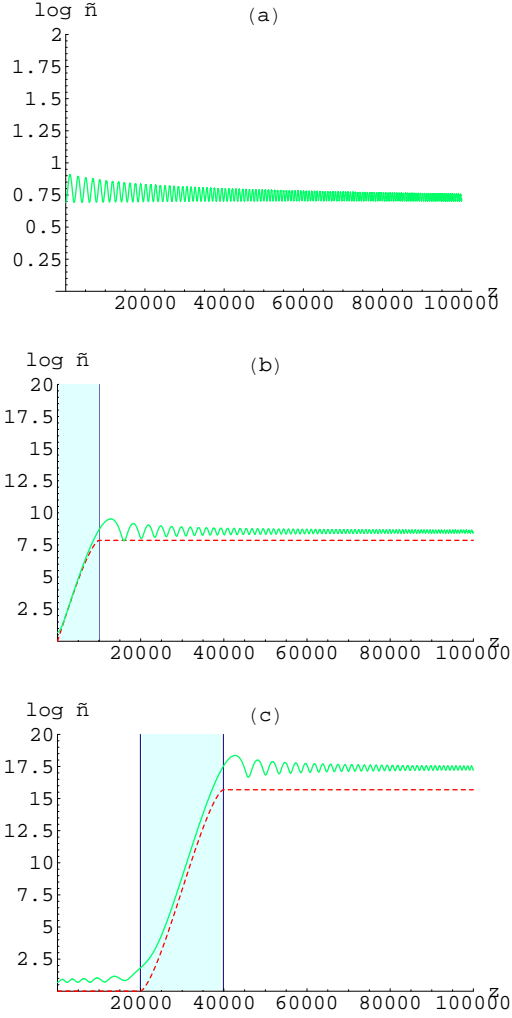}
		\caption{Numerical results obtained for the particle number with $|q_0|=10^{-3}$ and $\gamma=10^{-7}$. The plots correspond to (a) a type I mode, with $\beta=-3$, (b) a type II mode, with $\beta=0$ and (c) a type III mode with $\beta=3$. The solid line corresponds to the numerical solution in all three cases, while in (b) and (c) the dashed line gives the corresponding analytical prediction. The shaded area corresponds to the period that each mode spends inside the resonance band.}
	\label{fig:figure_5}
\end{figure}

Observing the plots shown in this figure, we see that all three particle-modes follow the expected behavior. The type I mode exhibits an oscillating particle number, with oscillations of decreasing amplitude and period, confirming that, in fact, no net particle number is produced\footnote{The positive value of $log\ \tilde{n}_k$ reflects the normalization chosen in Eq. (\ref{eq103}) and does not correspond to a physical particle number.}. 
Both the type II and the type III modes are exponentially amplified during a finite period, after which their particle number oscillates with decreasing amplitude and period, tending to an adiabatically constant value. Also, as expected, the type III mode exhibits an oscillating behavior before entering the resonance band. 

We can also conclude from Figure 5 that the particle number follows the predicted evolution inside the resonance band, the main differences between the numerical and the analytical solutions occurring near the endpoints of the resonance band, where the transitions between resonant and oscillating regimes take place. The discrepancies for low particle number are also due to the normalization of $\tilde{n}_k$, as discussed earlier. It is nevertheless clear that Eq. (\ref{eq100}) gives a quite good description of the resonant particle production regime in an expanding universe, although it underestimates the particle number density, and that the contribution of the non-resonant periods are subdominant, as expected.
 
The total particle number produced by the end of the resonant regime, $\tilde{n}_k(z_2)$, was computed for several values of $\beta$, keeping $|q_0|$ and $\gamma$ fixed at the values chosen above. The results we have obtained are shown in Figure 6.
 
\begin{figure}[htbp]
	\centering
		\includegraphics[scale=0.9]{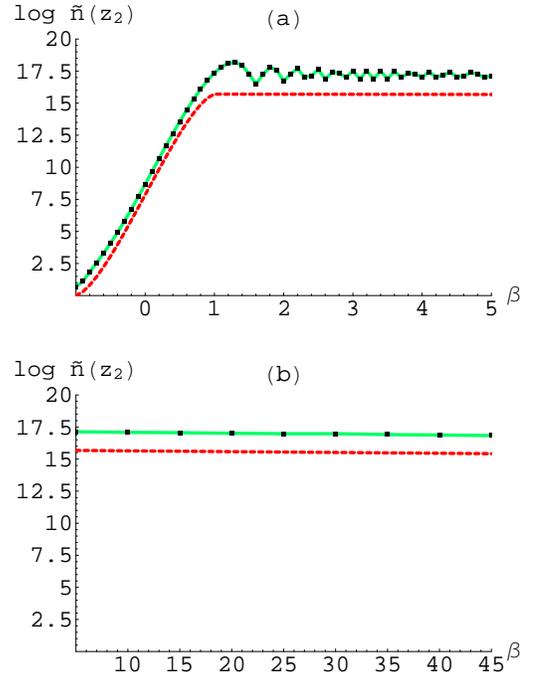}
		\label{fig:figure_6}
		\caption{Numerical results for the value of the total particle number produced by the resonance as a function of $\beta$. Plot (a) shows the results for small values of $\beta$, including type II and III modes, and plot (b) shows the results for large values of $\beta$, where only type III modes are present. The solid line corresponds to the numerical results while the dashed line gives the corresponding analytical prediction. All results correspond to $|q_0|=10^{-3}$ and $\gamma=10^{-7}$.}
\end{figure}
 
One observes that $\tilde{n}_k(z_2)$ increases with $\beta$ for type II modes and slowly decreases with $\beta$ for type III modes, in agreement with Eq. (\ref{eq101}). The main discrepancies between the numerical results and the analytical prediction are again due to the smooth transition between the resonant and non-resonant regimes at $z_2$, with the latter giving a subdominant contribution to the total particle number, which is more significant for type III modes. The different number of oscillations the modes undergo before entering the resonance band give the oscillations that can be observed in Figure 6. These are suppressed for large $\beta$ as the pre-resonance oscillations of the particle number give a negligible contribution for high momentum modes.

Numerical simulations also show that, as expected, the particle number in all modes decreases with $\gamma$, so that Hubble expansion may inhibit the resonant amplification of particle modes at early times.

Thus, we conclude that our analysis of the resonance regime gives a good description of the mechanism of particle production in branonium for an expanding universe and that the corrections arising from the non-resonant regimes are subdominant. Extrapolation of this discussion to high momentum modes is, however, difficult, as these will only be excited at late times, where the approximations we have considered may no longer hold. The production of high momentum particles is nevertheless suppressed, as we have concluded earlier. Furthermore, at each given time $z>0$, only the modes with $z_1<z$ have entered the resonance band, which means that only those modes for which
\beq \label{eq104}
A_{0k}<A_{0k}^{max}(z)\equiv{1+|q_0|(1+\xi z)\over 1-\gamma z}
\eeq
have started being excited. This gives the following momentum cut-off for modes which have already been amplified at $t>t_0$:
\beq \label{eq105}
k_{max}(t)={1\over2}\bigg[2\delta_0^2\bigg(1+6\alpha{\Delta t\over t_0}+\Omega_0^2\bigg)\bigg]^{1\over2}\bigg(1-20\alpha{\Delta t\over t_0}\bigg)^{-{1\over2}}~.
\eeq
Notice that this gives the expected endpoint of the resonance band in the limit $\alpha\rightarrow0$, given by Eq. (\ref{eq73}).

The generalization of these results for massive particles follows the same reasoning used in the non-expanding case, with the substitution $k_{phys}^2\rightarrow k_{phys}^2+m^2$ within our approximations. Note that, in the expanding case, the physical momentum of the particles is redshifted while their mass remains the same as the universe expands. This modification will then produce a more complex evolution of the parameter $A_k(z)$ in the modified Mathieu equation, Eq. (\ref{eq91}), given to lowest order by:
\beq \label{eq106}
A_k(z)=A_{k0}(1-\gamma z)+{4m^2\over\Omega_0^2}(1-\varsigma z)~,
\eeq
where $\varsigma\equiv{9\over10}\gamma$ and all other quantities are defined as before. Although this will alter the way the resonance band moves into higher momentum modes as the probe's orbit decays, the evolution of the modes is expected to follow the same qualitative behavior as in the massless case, so that we will not analyze this in more detail.

We, thus, conclude that, to lowest order, the parametric resonance survives in an expanding universe, although modes are excited only during a finite period. These results hold only at late times but we note once again that all quantities are measured with respect to the typical string values. We need, however, to bear in mind that Eq. (\ref{eq91}) is not the Mathieu equation and that several other terms in Eq. (\ref{eq52}) may become important at early times. These may alter the excitation time and the amplification of each mode, or even prevent any particle production. It is nevertheless clear that the effects of the expansion are suppressed at the typical energies involved in the probe's motion, so that one still expects a significant number of particles to be produced.


\subsection{Energy radiated into brane particle-modes}

As the resonance develops, the probe brane loses energy to excited $\eta$-particle modes and one may wonder whether the energy radiated by this mechanism is sufficient to affect the probe brane's motion.

Let us start by computing the energy density of massless particles produced by the resonance in a non-expanding universe. We have seen that each rescaled mode function $X_k$ has an associated harmonic oscillator energy function $E_k=\omega_k(n_k+1/2)$. For the mode functions $\chi_k=\sqrt{f}X_k$, the associated energy function is approximately the same in the non-relativistic and large distance limit, where $f\simeq 1$. Hence, the total energy density radiated into brane particle-modes is given by:
\beq \label{eq107}
\epsilon_p=\int {d^3k\over(2\pi)^3}E_k={1\over2\pi^2}\int_0^{+\infty}dk\ k^2\omega_k\bigg(n_k+{1\over2}\bigg)~,
\eeq
where we have used the momentum space isotropy discussed earlier in this work. After a significant number of particles has been produced, we may take $n_k\gg1/2$ and neglect the $1/2$ factor in Eq. (\ref{eq107}). Also, we know that particles are produced in a narrow resonance band centered at $k_c=\Omega/2$ with a band width $\Delta k_{res}\simeq \delta^2/\Omega=|q|\Omega/2$. The particle number distribution in momentum space can then be well approximated by a Gaussian distribution of the form:
\beq \label{eq108}
n_k(t)=n_{k_c}(t)\exp\bigg(-{1\over2}{(k-k_c)^2\over\Delta k(t)^2}\bigg)~.
\eeq
The gaussian distribution width $\Delta k$ should correspond to a fraction of the resonance band width, so that we may write $\Delta k=\lambda \Delta k_{res}$. The coefficient $\lambda$ is expected to be time-dependent as the central mode $k_c$ is more amplified than all the other modes in the resonance band, so that the distribution should become more sharply peaked about this value with time. The advantage of writing the particle number distribution in this form is that it allows us to further approximate the result by a $\delta$-function distribution, taking into account that $\Delta k/k_c\ll1$, as discussed earlier. Introducing the correct normalization, we may then write:
\beq \label{eq109}
n_k(t)\simeq n_{k_c}(t)\sqrt{2\pi}\Delta k(t)\delta(k-k_c)~.
\eeq 
Using $\omega_{k_c}\simeq k_c=\Omega/2$, we obtain:
\beqa \label{eq110}
\epsilon_p&\simeq& {\lambda\over\pi\sqrt{2\pi}}\bigg({\Omega\over2}\bigg)^3\Delta k_{res}n_{k_c}\nonumber\\
&\simeq& {\lambda\over\pi\sqrt{2\pi}}|q|\bigg({\Omega\over2}\bigg)^4 n_{k_c}~,
\eeqa
where the time dependence of $n_{k_c}$, $\lambda$ and consequently $\epsilon_p$ is implicit. Notice that, when $q\rightarrow0$, we have $\epsilon_p\rightarrow0$, as the resonance vanishes in this limit. To have an estimate of the order of magnitude of this quantity, take the example we have considered before in the discussion of particle production in a non-expanding universe, for which $|q|=10^{-4}$ and $\Omega=4\times10^{-5}$, in units of the string length. By solving the equations of motion numerically and fitting the obtained particle number distribution to the gaussian distribution given in Eq. (\ref{eq108}), we may determine the values of $\lambda$ and $n_{k_c}$. For an initial time $t_0=10^9$, we obtained the following results:
\beqa \label{eq111}
n_{k_c}(5t_0)&\simeq&7.5\times10^2~, \qquad \lambda(5t_0)\simeq 0.20~,\nonumber\\
n_{k_c}(10t_0)&\simeq&1.6\times10^7~, \qquad \lambda(10t_0)\simeq 0.12~.
\eeqa 

These two examples confirm that, indeed, the resonance width decreases in time. From Eq. (\ref{eq110}), we obtain for these two cases $\epsilon_p(5t_0)\simeq 3\times10^{-22}$ and $\epsilon(10t_0)\simeq4\times10^{-18}$, in units of the string energy scale.

The energy density radiated into brane particle-modes should then be compared with the energy density of the interbrane distance field $\phi$, which is given by $\epsilon=-\sigma/(2R)=-(\sigma\Omega/2)^{2\over3}$. For the example considered above, $\epsilon\simeq-2\times10^{-2}$, so that the value obtained for the energy density radiated into particles up to $t=10t_0=10^{10}$ is negligible when compared to the probe brane's energy density. In general, we obtain:
\beq \label{eq112}
{\epsilon_p\over|\epsilon|}\simeq{\lambda\over\pi\sqrt{2\pi}}{|q|\over\sigma^{2\over3}}\bigg({\Omega\over2}\bigg)^{10\over3} n_{k_c}~,
\eeq 
which gives the main result of this subsection. This means that, in the previous example, if we take $\lambda\sim0.01-0.1$, up to a $\eta$-particle number density of  $10^{23}-10^{24}$ can be produced without affecting significantly the motion of the probe brane.

These results can be easily generalized for the massive case, by taking into account the changes in $k_c$ and $\Delta k_{res}$ discussed earlier. For $\delta^2/2\ll k_c^2$, we obtain:
\beq \label{eq113}
{\epsilon_p\over|\epsilon|}\simeq {\lambda\over\pi\sqrt{2\pi}}{|q|\over\sigma^{2\over3}}k_c\bigg({\Omega\over2}\bigg)^{7\over3} n_{k_c}~,
\eeq
with $k_c=\sqrt{(\Omega/2)^2-m^2}$ and which, as expected, reduces to Eq. (\ref{eq112}) in the limit $m\rightarrow0$. As the momentum of the produced particles decreases with $m$, it is easy to conclude that the energy damped into brane particle-modes is smaller for more massive particles (recall that the strength of the resonance is not affected by the mass of the particles).

The analysis of the energy radiated into brane particle-modes is more difficult to perform in the expanding universe case, as more effects have to be taken into account. First, we need to recall the approximate relation between $\tilde{n}_k$ and the physical particle number. Next, we need to notice that the mode functions $\chi_k$ are redshifted by a factor of $a(t)^{-{3\over2}}$ with respect to the harmonic oscillator mode functions $X_k$. Finally, for the purposes of determining the energy density of the produced particles, we may take $\omega_k(t)\simeq k/a(t)$. Then, we have
\beq \label{eq114}
\epsilon_p(t)\simeq{1\over a(t)^5}\int {dk\ k^3\over8\pi^2}\tilde{n}_k~.
\eeq

We see that the energy density of the produced particles is redshifted by the usual power of $a^{-3}$, but is further reduced due to the redshift of the physical momentum and energy of the modes. We can use the expressions obtained for the particle number produced by the resonance in Eq. (\ref{eq101}) to determine $\tilde{n}_k$, as we have concluded that the effects of the non-resonant periods are subdominant. Writing them in terms of the comoving momentum of the modes, we conclude that, for type III modes, 
\beq \label{eq115}
\epsilon_p(t)\propto\int dk\ k^3\exp\bigg({q_0^2\Omega_0^2\pi\over8\gamma}{1\over k^2}\bigg)~.
\eeq

To determine the limits of this integral, one must recall that, at a given time, only a finite number of modes has entered the resonance band. In particular, one may use the value computed in Eq. (\ref{eq105}) as the upper limit of this integral, although some of these modes have not yet been completely excited by the resonance. The lower limit of the integral will be the first type III mode, with $k=k_*\equiv k_{c0}\sqrt{1+|q_0|}$. This gives a finite energy density radiated into brane modes at each finite time $t>t_0$. This energy density is redshifted by Hubble expansion, but at the same time modes with increasing energy enter the resonance band. This process will, however, stop at some point, as our approximations will break down when the motion of the probe becomes relativistic and gets too close to the central stack. Even if the resonance mechanism persists in this limit, although certainly in a different form, the probe will eventually modify its trajectory, either by colliding with the central stack or via some other mechanism that stabilizes its motion. 

The contribution of type II modes is obviously finite and can be computed by integrating Eq. (\ref{eq114}) using the corresponding expression for $\tilde{n}_k$ given in Eq. (\ref{eq101}), with a lower limit $k_{c0}$ and an upper limit $k_*$ for the integral. No particular insight is, however, gained by computing the exact expressions for the energy density for both type II and type III modes. Instead, we note that the dominant contribution to the energy density will be given by high momentum modes which enter the resonance band at late times. For these, the effects of the expansion of the universe will be less significant and one may use the results obtained in the non-expanding case to compute the energy density produced by the resonance. In this case one should use the values of $q$ and $\Omega$ at the end of the resonance mechanism.


\subsection{Radiation into bulk modes}

As discussed in \cite{Burgess}, the probe's trajectory in the background created by the central stack may induce not only the production of brane-bound particles but also radiation into bulk modes, namely gravitational, RR-form and dilatonic fields. This effect is due to the accelerated motion of the probe brane and most of the power is radiated into lower-spin fields, in particular scalar fields which couple to the brane's orbital monopole moment.

The probe will thus lose energy through this process, contributing to the decay of its orbit. This will certainly modify the mechanism of resonant production of particle-modes in the brane, namely by varying the orbital frequency, which determines the comoving momentum of the modes that are excited by the parametric resonance. One then expects the effects of radiation into bulk modes to be quite similar to those induced by the universe's expansion, so that we will not analyze this process in detail. Nevertheless, it is important to estimate its contribution to the decay of the probe's orbit.

The power radiated into bulk scalar modes was estimated in \cite{Burgess}, where it was shown that the number of orbits the probe can complete before the interbrane distance becomes of order $l_s$ is approximately given by:
\beq \label{eq116}
\mathcal{N}\simeq{2\over3\pi}\bigg({r_i\over l_s}\bigg)^{3\over2}{1\over(g_s^3N)^{1\over2}}~,
\eeq
where $r_i$ is the initial value of the physical radius of the orbit, assuming it is circular. (This gives a good estimate for the decay time, even though we are interested in small eccentricity orbits for the resonance process. $\mathcal{N}$ corresponds to the ratio of the decay time to the initial orbital period.) This expression shows that, for the probe to complete a large number of orbits before decay, it needs to be at a large distance from the central stack, in units of the string length. Also, one needs $g_s^3N\ll1$ to obtain a sufficiently large value of $\mathcal{N}$. 

This value should be compared to the number of orbits of the probe necessary for the resonance to be effective, which in the non-expanding case is of order $(2\pi\mu_k)^{-1}\sim(\pi|q|)^{-1}$. Hence, the production of a significant number of particles requires $\mathcal{N}\gg(\pi|q|)^{-1}$, which gives for the initial radius of the orbit:
\beq \label{eq117}
{r_i\over l_s}\gg\bigg({3\over2|q|}\bigg)^{2\over3}N^{1\over3}g_s~.
\eeq
For example, if $|q|\simeq10^{-3}$ and there are 10 branes in the central stack, we need the initial interbrane distance to be larger than about $300g_sl_s$ for the resonance to be effective, which is not too large a number taking into account that $g_s$ is parametrically small. 

As, in the analysis of the parametric resonance mechanism, we have assumed the interbrane distance to be large compared to the string length, we expect the effects of radiation into bulk modes to be initially negligible. This process should, however, become more important as the orbit decays and increases its acceleration. It is even possible that, at late times, it overcomes Hubble expansion as the main energy loss process. Nevertheless, the end of the resonance should still be determined by the break down of the non-relativistic and large distance approximation, as this also controls the amount of energy damped into bulk modes.


\section{Effects of transverse space compactification}

So far we have considered the probe to move at distances from the central stack which are small compared to the typical size of the transverse space directions, so that we may neglect the effects of compactification on the probe's motion. In this section, we will analyze the leading order effects introduced by the finite size of these directions, and show that they may lead to the creation of orbital angular momentum, a necessary condition for resonant particle production to take place. 

We will consider the simplest case of compactifying the three transverse directions on an isotropic 3-torus of size $R_{\perp}$, which is defined by the identifications $y^i\leftrightarrow y^i+R_{\perp}$. As mentioned earlier, the harmonic function associated with the central stack configuration needs to be modified in this case by including the appropriate ``brane images", according to Eq. (\ref{eq10}). These images are placed at points with coordinates $y^i=n_iR_{\perp}$, with integer $n_i$, defining a hypercubic lattice corresponding to the covering space of the 3-torus, $(\textbf{R}/\textbf{Z})^3$. Although toroidal compactifications are too simple to give realistic particle physics phenomenology, this will be sufficient to illustrate the main effects of compactification on the probe brane's trajectory.

The interbrane potential associated with the generalized harmonic function Eq. (\ref{eq10}) is obtained by computing the propagator for a massless field on a torus, which satisfies:
\beq \label{eq118}
\nabla^2G(\mathbf{y},\mathbf{y'})=\delta(\mathbf{y}-\mathbf{y'})-{1\over V_{\perp}}~,
\eeq
where $V_{\perp}\equiv R_{\perp}^3$ is the volume of the transverse 3-torus. The term $-1/ V_{\perp}$ is included for consistency, so that the integral over the compact manifold of both sides of Eq. (\ref{eq118}) vanishes. If one expands the Green's function $G(\mathbf{y},\mathbf{y'})$ in terms of eigenfunctions of the Laplacian operator $\nabla^2$, one concludes that the $-1/ V_{\perp}$ term removes the unphysical zero-mode which makes the Green's function diverge. This term arises naturally from the curvature of the non-compact 4-dimensional spacetime \cite{Kachru}. It has been shown that one can write the massless propagator on the 3-torus in terms of an integral involving the Jacobi-theta function $\theta_3$ \cite{Martineau}:
\beq \label{eq119}
G(\mathbf{y})\equiv G(\mathbf{y},\mathbf{0})={1\over R_{\perp}}\int_0^{\infty}\mathrm{d}s\ \Big[1-\prod_{i=1}^3\theta_3\bigg({\pi ~y^i\over R_{\perp}},e^{-4\pi^2s}\bigg)\Big]~.
\eeq

Although this expression gives the full compact space propagator, we are mainly interested in the leading order modifications introduced by compactification when the probe is moving closer to the central stack than to any of its images (orbits around any of the brane images are equivalent due to the symmetries of the hypercubic lattice). In \cite{Shandera}, an expansion of this propagator about the origin was computed using Ewald's method for calculating potentials in hypercubic lattices in the context of solid state physics. The leading corrections to the central $1/r$ potential are given by:
\beq \label{eq120}
G(\mathbf{y})=-{1\over 4\pi r}-{r^2\over 6V_{\perp}}-{C_S\over V_{\perp}^{1\over3}}-A_4h_4(\mathbf{y})-A_6h_6(\mathbf{y})-\ldots
\eeq
The values for the coefficients of the various terms were found numerically to be $C_S=-0.21$, $A_4=0.44$ and $A_6=0.0072$. The functions $h_4(\mathbf{y})$ and $h_6(\mathbf{y})$ are harmonic functions of order $(r/ R_{\perp})^4$  and $(r/ R_{\perp})^6$, respectively, and the series continues with harmonic terms of higher (even) orders. The leading harmonic correction to the propagator is given by:
\beq \label{eq121}
h_4(\mathbf{y})={1\over R_{\perp}^5}\Bigg[\sum_{i=1}^3(y^i)^4-3\sum_{i\neq j=1}^3(y^i)^2(y^j)^2\Bigg]~.
\eeq

Recalling that the function $h(\mathbf{y})$ contributes to the interbrane potential through the graviton-dilaton and RR-form interactions, we may write this potential for $Q_6\ll r \ll R_{\perp}$, to order $(r/R_{\perp})^4$ and discarding constant terms, as
\beqa \label{eq122}
V(\mathbf{y})&=&-2Q_6\bigg[{1\over r}+{2\pi\over 3R_{\perp}}\bigg({r\over R_{\perp}}\bigg)^2+\nonumber\\
&\ &+{4\pi A_4\over R_{\perp}}\bigg({r\over R_{\perp}}\bigg)^4f\bigg({y^i\over r}\bigg)+\ldots\bigg]~,
\eeqa
where we defined the function
\beq \label{eq123}
f\bigg({y^i\over r}\bigg)\equiv\Bigg[\sum_{i=1}^3\bigg({y^i\over r}\bigg)^4-3\sum_{i\neq j=1}^3\bigg({y^i\over r}\bigg)^2\bigg({y^j\over r}\bigg)^2\Bigg]~.
\eeq

We conclude that, to this order, the interbrane potential is modified by two terms. Both terms correspond to repulsive contributions to the potential arising from the overall attraction of the image branes and their coefficients are suppressed by a factor $1/R_{\perp}$, so that their effects are negligible for large transverse volume. The ``jellium" term, of order $(r/R_{\perp})^2$ and whose name arises in the context of solid state physics, gives an isotropic contribution while the ``asymmetry" term, of order $(r/R_{\perp})^4$, gives an anisotropic contribution to the potential.

We will be more interested in the effects of the asymmetry term, as its breaks the rotational symmetry of the transverse space and leads to the generation of orbital angular momentum. The effects of the jellium term will, however, be more significant close to the central stack. It is easy to show that, for large transverse volume, this term will, to leading order, make the probe brane's orbits precess. This is due to its repulsive nature and, for small eccentricity orbits, the deficit angle is approximately given by $\Delta\theta\simeq32\pi^2(l^2/R_{\perp}Q_6(T_6V_3)^2)^3$. This effect will, however, be suppressed at late times as the probe's angular momentum is redshifted away by Hubble expansion.  

The inclusion of the asymmetry term makes the analysis quite difficult, as the potential is no longer central and depends on both angular coordinates in the transverse space. This may lead, in particular, to non-planar orbits of the probe brane. However, if we set $y^3=\dot{y}^3=0$ initially, the probe will feel no force along this direction and its trajectory will be confined to the $(y^1,y^2)$ plane as before. We will focus on this particular case, bearing in mind that in general non-planar trajectories may arise. 

Within these assumptions, the problem reduces, as before, to the evolution of the canonically normalized complex scalar field $\phi$, defined in Eq. (\ref{eq17}), in an expanding universe. Its potential can be written as
\beq \label{eq124}
V(\phi)=-{\sigma\over|\phi|}-\sigma_J|\phi|^2-\sigma_A(\phi_R^4+\phi_I^4-6\phi_R^2\phi_I^2)~,
\eeq
where $\phi_R$ and $\phi_I$ are, respectively, the real and imaginary parts of the field, $\sigma$ was defined in Eq. (\ref{eq27}) and
\beqa \label{eq125}
\sigma_J\equiv{8\pi\over3}{Q_6\over R_{\perp}^3}~,\qquad \sigma_A={32\pi A_4\over T_6V_3}{Q_6\over R_{\perp}^5}
\eeqa
give the strength of the jellium and asymmetry terms. This potential is no longer invariant under the global U(1) symmetry of the non-compact case, signaling that the probe's angular momentum is no longer conserved. This is quite similar to the U(1)-violating potential governing the evolution of the scalar field in the Affleck-Dine mechanism \cite{Affleck}, where non-conservation of angular momentum plays a crucial role in generating the U(1) baryon number asymmetry in our universe. The potential remains, however, invariant under the hypercubic group symmetries $\phi_{R,I}\rightarrow-\phi_{R,I}$ and $\phi_R\leftrightarrow\phi_I$. In terms of the polar angle $\theta$, this means that $\pi/2$ rotations as well as reflections about the $\theta=\pi/4$ axis are preserved by the asymmetry term, as one can easily conclude by writing Eq. (\ref{eq123}) for $y^3=0$ as
\beq \label{eq126}
f(\theta)=1-2\sin^2(2\theta)~.
\eeq  

Hence, we only need to consider initial conditions such that $\phi_R\geq\phi_I\geq0$ or equivalently $0\leq\theta\leq\pi/4$. The U(1) symmetry violation precludes a complete analytical description of the orbits. However, it is not difficult to obtain a qualitative insight on the main features of the probe's trajectories. If the probe is placed significantly far from the central stack (but still closer to it than to any of the image branes), the asymmetry term will not be negligible and some angular momentum will be created or destroyed. If the probe has no angular momentum initially, as we would expect immediately after inflation, then it will necessarily acquire some and be placed in an orbit around the central stack, instead of just falling towards it along the radial direction. As both the jellium and the asymmetry interactions are repulsive, it is possible that this trajectory does not remain bound to the central stack, becoming connected to one or more branes in the hypercubic lattice. Hubble expansion will, however, redshift the probe's energy and angular momentum, so that one expects the probe to become bound to only one brane stack at late times. As the orbital radius decreases, the U(1)-violating term becomes less significant and the particle number associated with $\phi$ should asymptotically become constant, i.e. the angular momentum should vary only due to the universe's expansion. One also expects the jellium term, as well as the asymmetry term, to induce some precession of the orbital axis, which should stop at late times when these terms become negligible.

Due to the hypercubic symmetries, along $\theta=0,\pi/4$ the force acting on the probe will be in the radial direction, as $f'(\pi/4)=f'(0)=0$, and no angular momentum will be created in these particular directions. For $0<\theta<\pi/4$, angular momentum creation should, however, be a generic feature.

To have a better understanding of how the initial conditions affect the amount of angular momentum created or destroyed, we have computed the force acting on the probe brane and the associated torque, given by:
\beqa \label{eq127}
\mathbf{F}=-\mathbf\nabla V=-{\partial V\over\partial\phi_R}\mathbf{e_R}-{\partial V\over\partial\phi_I}\mathbf{e_I}~,\nonumber\\
\tau=|\mathbf{\phi}\times\mathbf{F}|=-\phi_R{\partial V\over\partial\phi_I}+\phi_I{\partial V\over\partial\phi_R}~,
\eeqa
where we have considered the field $\phi$ as a vector in the $(\phi_R,\phi_I)$ plane. Recall that the torque gives the variation of the angular momentum, which in terms of the particle number density $n$ defined in Eq. (\ref{eq20}) can be written as:
\beq \label{eq128}
{dn\over dt}+3Hn=\tau~.
\eeq

We have plotted the isotorque contours for some particular parameter values in Figure 7. We have restricted the position of the probe to the region $y^1,y^2<0.5R_{\perp}$, where the expansion of the potential about $r=0$ is sufficiently accurate and the probe's motion is bound to the central stack (near the limits of this region, the orbits may not be bound to the central stack, as the actual region where the force points towards the central stack has a more complicated shape).

\begin{figure}
	\centering
		\includegraphics[scale=0.6]{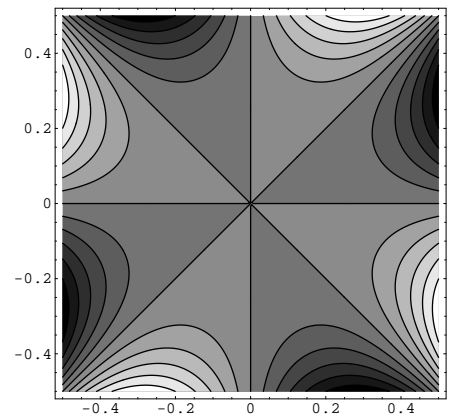}
	\label{fig:figure_7}
	\caption{The torque acting on the probe brane in the ($y^1,y^2$) plane. Coordinates are normalized to the size of the transverse torus, with the value $R_{\perp}=2\times10^4$ in this example, in units of the string length. We considered $Q_6=100$ and $T_6V_3=1$ in these units. Note that the torque changes sign when crossing one of the $\theta=n\pi/4$ axis, with integer $n$, being negative for $0<\theta<\pi/4$. Also, the absolute value of the torque increases as one moves away from the origin.}
\end{figure}

Observing this figure we see that the torque is maximized close to the boundary of the region mentioned above (darker areas for $0\leq\theta\leq\pi/4$). Also, the longer the probe's trajectory remains close to this boundary the more angular momentum it is likely to gain. Thus, maximum angular momentum creation should occur if the probe is placed below the centre of the hypercubic cell.

The initial value of the Hubble parameter $H$ will also affect the amount of angular momentum created. If it is of the same order of magnitude as the orbital frequency, one expects the probe to be driven into lower torque regions within a few periods. Lower values of the Hubble parameter should allow the probe's angular momentum to oscillate significantly as the probe moves through alternate regions of positive and negative torque, but no net angular momentum will be gained after completing the first few orbits. The asymptotic value of the comoving particle number $N=na^3$ will, hence, be determined by the initial path of the probe.

We have simulated the evolution of the probe brane numerically to illustrate this discussion. The initial values for the real and imaginary parts of the field were defined as follows:
\beq \label{eq129}
\phi_{R,I}(t_0)=\sqrt{T_6V_3\over2}\delta_{R,I}R_{\perp}~, \qquad \dot{\phi}_{R,I}=0~,
\eeq
with $\delta_R>\delta_I$, giving $0<\theta<\pi/4$, and $\delta_{R,I}<0.5$, which ensures the probe is placed initially within the region where it is bound to the central stack. We have also computed the eccentricity of the orbit, as defined in Eq. (\ref{eq29}), with the probe's energy density including both the jellium and asymmetry terms. This quantity should only become meaningful at late times, when the latter terms become negligible, but its evolution tracks the creation of angular momentum along the probe's trajectory. 

Figure 8 shows an example of the results obtained for the probe's orbit. In this case, the probe is placed below the centre of the hypercubic cell and, as expected, a significant angular momentum is produced during its motion. One observes, as expected, a small precession of the orbital axis and the decay of the orbit into regions of lower torque, which drives the orbital eccentricity towards a constant value. Initial conditions were chosen in this example so that the inverse Hubble parameter is of the same order of magnitude as the orbital period and, hence, the orbit is quickly stabilized. The eccentricity tends in this example to a constant value of $e\simeq0.28$. Notice the similarities with the example illustrated in Figure 1, the main difference residing in the fact that, in this case, the probe has initially no angular momentum.

\begin{figure}
	\centering
		\includegraphics[scale=0.6]{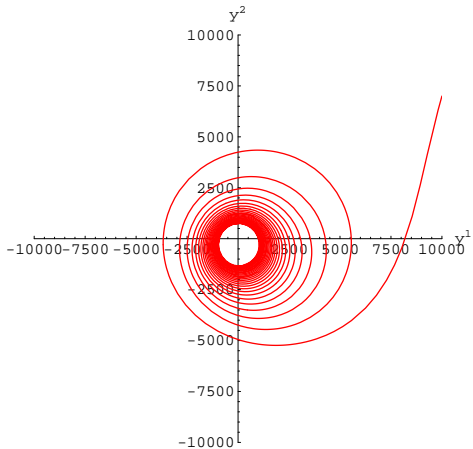}
	\label{fig:figure_8}
	\caption{Results obtained numerically for the probe brane's trajectory for $Q_6=100$, $R_{\perp}=2\times10^4$ and $T_6=V_3^{-1}=(2\pi)^{6}$. The probe is initially stationary at the point $(\delta_R,\delta_I)=(0.50,0.35)$ for $t_0=2\times10^6$ and we consider a matter-dominated universe with $\alpha=2/3$. The asymptotic value obtained for the eccentricity is $e\simeq0.28$.}
	\end{figure}

In Figure 9 we plot the asymptotic values of the eccentricity obtained numerically for different initial positions in the plane $(\phi_R,\phi_I)$, considering two distinct values of the initial Hubble parameter in a matter-dominated universe. In both cases, one observes that smaller eccentricity orbits are obtained near the larger torque regions below the centre of the hypercubic cell, as expected. We see that, in general, some angular momentum is produced, although the majority of initial conditions leads to highly eccentric orbits, with $e>0.9$. We also conclude that the probe's orbit is generically less eccentric if its motion starts at later times, so that the initial Hubble damping of the asymmetry term is smaller. As one may observe in Figure 9, in particular in plot (b), the combination of the asymmetry term and Hubble expansion makes the final eccentricity vary in a non-trivial way with the initial position of the probe, namely near $\delta_R\simeq0.5$, where the orbits may in some cases become unbound. Note that these results correspond to particular values of the parameters of the problem and that the values of the asymptotic eccentricity are highly dependent on these parameters, namely the relevant distance scales $Q_6$ and $R_{\perp}$.

\begin{figure}
	\centering
		\includegraphics[scale=0.77]{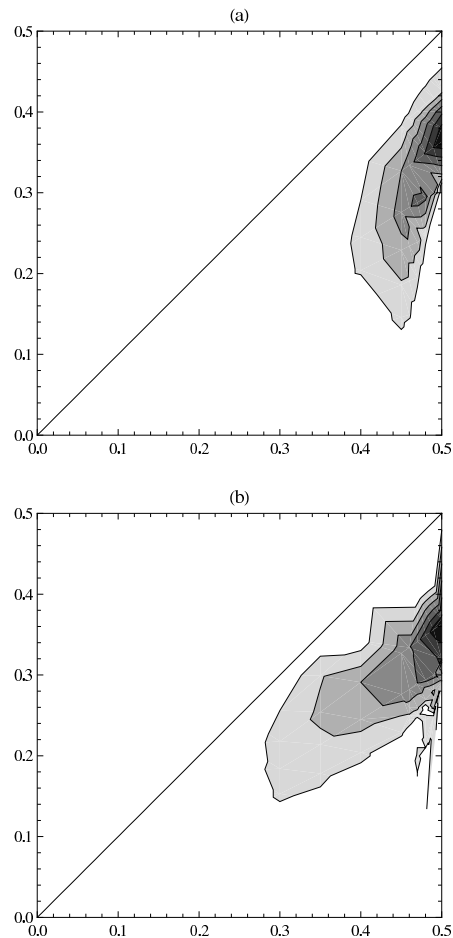}
	\label{fig:figure_9}
	\caption{Contour plots obtained numerically for the asymptotic value of the orbital eccentricity in the plane $(\delta_R,\delta_I)$, with $\delta_R\geq\delta_I$ (with similar results for $\delta_R\leq\delta_I$), assuming $Q_6=100$, $R_{\perp}=2\times10^4$ and $T_6=V_3^{-1}=(2\pi)^{6}$. Both plots were obtained for a matter-dominated universe with $\alpha=2/3$ and the motion of the probe starts at (a) $t_0=1\times10^6$ and (b) $t_0=5\times10^6$. All parameters are given in units of the string length. Contours were plotted for $e=n/10$, with $n=3,\ldots,\!9$, and darker regions correspond to lower eccentricities.}
\end{figure}

These results show that a significant amount of angular momentum can be created by U(1)-violating terms in the probe's potential arising from compactification of the transverse dimensions, which is a key ingredient for the development of the parametric resonance we have studied in this work. The simple 3-torus compactification thus ilustrates how low eccentricity orbits can be created for a range of initial conditions, even if the probe has initially no angular momentum.


\section{Stabilization and cosmological implications}

Potential applications of the resonant particle production mechanism to cosmology will be largely determined by the system's final configuration. As we have discussed before, the $D6-\overline{D6}$ branonium we have analyzed in this work is unstable, as the probe (anti)brane loses energy and angular momentum due to Hubble expansion. At some stage, the distance between the probe and the central stack of branes will consequently become of the order of the string length (although our approximations will break down long before this happens). When this occurs, the scalar mode associated with open strings stretching between the probe and the source branes becomes tachyonic, signaling an instability of the system \cite{Sen}. Condensation of this tachyonic mode will then lead to the annihilation of the probe brane with one of the branes in the central stack. The non-supersymmetric nature of the $D6-\overline{D6}$ configuration, which is a general property of all systems mixing brane and antibrane states, is behind this instability. Namely, it makes the system evolve into a $1/2$ BPS state with $(N-1)$ parallel $D6$ branes, with the same total charge as the original configuration but with a lower energy. 

Interesting scenarios may arise if some additional mechanism stabilizes the probe brane at a finite distance from the central stack, in which case the produced particles may survive after the resonance mechanism ends. An attractive possibility for stabilizing multiple brane systems was suggested in \cite{Dvali}. When supersymmetry breaking occurs, one or more fields within the supermultiplets of the theory typically acquire masses, their value being set by the energy scale at which SUSY is broken. Massive gauge potentials will then induce short-range interactions which, if repulsive, may balance the gravitational atraction between different branes. Notice that the presence of a probe antibrane itself breaks SUSY but that this does not induce bulk field masses, so that other sources of SUSY breaking need to be considered in this case. 

Suppose, for example, that the $C_1$ RR-form and consequently its magnetic dual form $C_7$ become massive after SUSY breaking, the same happening to the dilaton field, $\Phi$. Both the RR-form and dilaton-mediated interactions will then be described by Yukawa potentials whose range is determined by the masses $m_{RR}$ and $m_{\Phi}$, respectively. If $m_{RR}\ll m_{\Phi}$, the dilaton-mediated interaction will be exponentially suppressed for interbrane distances of order $m_{RR}^{-1}$. Assuming $Q_6\ll m_{RR}^{-1}\ll R_{\perp}$, we may write the relevant terms in the potential for $r\sim m_{RR}^{-1}$ as

\beq \label{eq130}
V(r)=M^4\bigg(\lambda {e^{-m_{RR}r}\over r}-{1\over r}\bigg)~,
\eeq        
where $M^4$ gives the overall constant factor. The constant $\lambda$ parametrizes the effective strength of the RR-interaction compared to the gravitational atraction and one expects $|\lambda|>1$ as the dilaton-mediated interaction is negligible. If, as we have assumed so far, the probe is a $\overline{D6}$-brane, the RR-interaction, although short-ranged, is attractive and cannot balance the $1/r$ gravitational part. If, however, the probe is a $D6$-brane, we have $\lambda>0$ and stabilization of the system may be possible. In this case, for $r\gg m_{RR}^{-1}$, the RR-interaction is exponentially suppressed and the probe's motion is governed only by the gravitational term. The evolution of the probe brane will, in this case, be very similar to that of a probe antibrane in a massless RR-form background which we have studied in this work, the potential having a smaller strength than in the latter case. This difference does not modify the qualitative features of the probe's trajectory and all the results we have derived in this work remain valid using an effective central stack charge $Q_6^{eff}<Q_6$. In particular, one expects the parametric resonance to develop in the probe's world-volume under the same conditions as before and the effects of transverse space compactification to provide the necessary source of angular momentum for similar initial conditions.

The fate of a probe $D6$-brane will, however, be quite different than that of a probe $\overline{D6}$-brane. As the probe's orbit decays due to the universe's expansion, the interbrane distance will eventually become of order $m_{RR}^{-1}$. At these distances, the repulsive RR-interaction becomes relevant to the probe's motion and stabilization may be possible. The potential given in Eq. ({\ref{eq130}}) has a local minimum formally given by:
\beq \label{eq131}
r_0=-m_{RR}^{-1}\bigg[1+W\bigg(-{1\over\lambda e}\bigg)\bigg]~,
\eeq 
where $W(z)$ is the Lambert W-function, defined as the inverse of $f(W)=We^{W}$. After its angular momentum has been redshifted away by Hubble expansion, one expects the probe to settle at this local minimum. This may not be the absolute minimum of the full potential, which includes the dilaton-mediated interaction and other supergravity and string theory corrections closer to the central stack. Hence, absolute stabilization of the probe cannot be guaranteed, but it is reasonable to expect at least a long-lived metastable state.

From the properties of the Lambert W-function, one concludes that $r_0=0$ for $\lambda=1$, being positive for $\lambda>1$, and that $r_0$ strictly grows with $\lambda$. Hence, as expected, stabilization at large distances from the central stack is only possible if the RR-repulsion is stronger than the gravitational attraction. In this case, one expects the probe brane to stabilize away from the central stack at a distance $r_0\sim m_{RR}^{-1}$.

The scale of supersymmetry breaking will then determine the cosmological implications of the particle production mechanism analyzed in this work. If SUSY is broken at very high energies, inducing a large mass for the RR-potential (but still smaller than the dilaton's mass), the probe will stabilize very close to the central stack and no tachyonic modes are present to induce annihilation. The stable probe and the central stack will then be characterized by a general broken gauge symmetry group $U(N)\times U(M)$, where $M\ll N$ is the number of branes in the probe, with an exponentially large number of particles charged under the $U(M)$ gauge group if the resonance develops before stabilization occurs. This may be relevant, for example, to the generation of the baryon asymmetry in our universe. On the other hand, a parametrically small soft SUSY breaking mass will stabilize the probe at large distances. The parametric resonance may then be relevant for dark matter production, as particles in the probe's world-volume will necessarily interact weakly with the visible sector if the latter is embedded in the central stack. As we have shown, these particles may be supermassive (although parametrically small compared to the string scale), so that this scenario could provide an effective mechanism for a non-thermal production of WIMPZILLAS \cite{Kolb}.

One also expects the parametric resonance to amplify other bosonic modes living in the probe's world-volume, such as Yang-Mills fields if the probe has more than one brane. If these mediate baryon number-violating forces, their interactions with central stack fields could then provide a mechanism for baryogenesis as the one discussed in \cite{Linde}, in which case the probe would need to be stabilized before nucleosynthesis.

It is also possible for fermionic particles living in the probe's world-volume to be produced in resonance, although Pauli blocking makes this process significantly different from the bosonic case, according to the discussion given in \cite{Greene}. Thus, if the probe can be stabilized, a resonant production of fermionic dark matter particles may also be achieved through this process.

One also expects the interbrane distance field to oscillate about the local minimum of the potential, possibly generating a second stage of resonant amplification. This may be induced from coupling to other fields, as in the standard preheating mechanism, or gravitationally, as in the case described in this work. Although this second stage may contribute significantly to the final particle number density in the probe brane's world-volume, we will not discuss them in further detail but rather refer the reader to the discussions given in this work and in the literature \cite{preheating, Kofman, otherbranepreheating}.

We emphasize that all these scenarios depend on the particular embedding of the Standard Model fields in this setup, an issue for which there is still no complete answer and that is closely related to the nature of SUSY breaking and its communication to the Standard Model.


\section{Conclusions}

In this work we have analyzed the mechanism of particle production in $D6-\overline{D6}$ branonium systems, as well as in $D6-D6$ systems with SUSY breaking. We have shown that, in the limit of large distances and small velocities and for small eccentricity orbits, a parametric resonance will develop, producing scalar particles confined to the probe brane's world-volume. Massless particles are produced in a narrow resonance band centered at the comoving momentum corresponding to one half of the probe's orbital frequency $\Omega$, as if resulting from the direct decay of the particles associated with the interbrane distance field. The strength of the resonance is given by the exponent $\mu_k$ defined in Eq. (\ref{eq72}). Massive particles are produced in the same way, with energies of order $\Omega/2$. These energies are small compared to the string energy scale but, as the latter should be close to the Planck scale, very massive particles can be produced by the resonance mechanism, according to Eq. (\ref{eq84}). For both massless and massive particles, the associated particle number grows exponentially in time and a large number of particles can be produced by this mechanism, just as in the case of preheating after inflation \cite{preheating,Kofman}. However, the resonance surely requires the probe to complete a large number of orbits before the particle-modes can be significantly amplified, due to the small value of the exponent $\mu_k\simeq{q\over2}\ll1$.

Hubble expansion of the 4-dimensional effective flat FRW spacetime makes the probe's orbit decay, redshifting its energy and angular momentum. This alters the development of the parametric resonance significantly if the probe's motion does not occur at sufficiently late times. If, however, its motion begins late in the history of the universe, at least compared to typical string times, the resonance will still develop with the associated resonance band being shifted towards higher momentum modes. Each mode will be excited during the finite period it spends inside the resonance band, after which the associated particle number becomes an adiabatic invariant. A significant number of particles, given appproximately by Eq. (\ref{eq101}), can be produced if this period is sufficiently long for the resonance to develop, according to the condition $q_0^2\gsim\gamma$, where $\gamma$ quantifies the effects of the expansion and was defined in Eq. (\ref{eq93}). 

We have also concluded that a large number of brane-bound particles can be produced without affecting the probe's orbital motion, as can be seen in Eq. (\ref{eq112}). Radiation into bulk closed string modes will also have a negligible effect if the probe moves at sufficiently large distances, as assumed for the validity of our study.

A realistic implementation of this mechanism requires particular initial configurations with large interbrane distances, non-relativistic velocities and almost circular orbits. We have shown that angular momentum creation may result from the effects of compactification of the directions transverse to the branes, illustrating this for the case of a compact 3-torus, where low eccentricity orbits can be produced for some range of initial conditions.

In this work, we have also discussed how, for a probe $D6$-brane, an interbrane potential may arise from supersymmetry breaking, if e.g. the relevant RR-form and the dilaton, $\Phi$, gain a mass in the process. Stabilization of the probe is then possible at late times and at distances of order $m_{RR}^{-1}$, if $m_{\Phi}\gg m_{RR}$. Such a stabilization avoids annihilation with the central stack, as necessarily happens for a probe antibrane, making the parametric resonance regime potentially relevant for baryon number or dark matter generation, for example. 

In our analysis, the parametric resonance arises only for $D6-\overline{D6}$ or $D6-D6$ systems, where the $1/r$ potential allows the probe to move in closed elliptical orbits in a non-expanding universe. It is crucial for the development of the resonance that the interbrane distance field $\rho$ exhibits an oscillating behavior, although the orbits do not need to close, as we have seen for the case of an expanding universe. Thus, other $Dp-\overline{Dp}$ or $Dp-Dp$ branonium systems with $p<6$, where the probe trajectories are necessarily unbounded from below, cannot exhibit such resonant particle production. Additional stabilization potentials may, however, provide the required oscillatory behavior, so that it may be possible to find a resonant particle production mechanism in other cases.

It is nevertheless clear that resonant particle production in branonium systems may play an important role in our universe's evolution and we hope with this work to motivate future exploitation of their properties and applications.

\vskip 0.2cm


\centerline{\bf {Acknowledgements}}
\vskip 0.2cm
The authors would like to thank Graham Ross, Martin Schvellinger, Andre Lukas and Yang-Hui He for useful discussions. JGR is supported by FCT (Portugal) under the grant SFRH/BD/23036/2005. JMR is partially supported by the EC Network $6^{th}$ Framework Programme Research and Training Network ``Quest for Unification" (MRTN-CT-2004-503369). This work was partially supported by the EU FP6 Marie Curie Research and Training Network ``UniverseNet" (MRTN-CT-2006-035863). 

\vskip 0.5cm

\vfill

\appendix
\section{Computation of the particle number for $H>0$}

The particle number produced by the parametric resonance can be estimated by computing the integral in Eq. (\ref{eq100}). This gives the physical particle number apart from constant factors, subdominant non-resonant contributions and also an $a^{-3}$ redshift factor due to Hubble expansion. We have,

\beq \label{eqA1}
\log n_k^{res}(z)=\left\{
\begin{array}{l}
0,\qquad\qquad\qquad\ z<z_i~,\\
2\int_{z_i}^{z}\mu_k(z')dz',\ \ z_i\leq z\leq z_2~,\\
2\int_{z_i}^{z_2}\mu_k(z')dz',\ \ z> z_2~.
\end{array}\right.
\eeq

In order to compute this function, let us write the exponent $\mu_k(z)$ in the form:
\beq \label{eqA2}
\mu_k(z)={1\over2}\sqrt{az^2+bz+c}~,
\eeq
where 
\beqa \label{eqA3}
a&=&q_0^2\xi^2-A_{k0}^2\gamma^2~,\nonumber\\
b&=&2(q_0^2\xi+A_{k0}(A_{k0}-1)\gamma)~,\nonumber\\
c&=&q_0^2-(A_{k0}-1)^2~.
\eeqa
For all modes of interest, $a<0$, while we have $c\geq 0$ for type II modes and $c<0$ for type III modes. The sign of $b$ will not affect our results. It is also useful to define $d^2\equiv b^2-4ac$, which is explicitly given by
\beq \label{eqA4}
d^2=4q_0^2\big(2A_{k0}(A_{k0}-1)\xi\gamma+(A_{k0}-1)^2\xi^2+A_{k0}^2\gamma^2\big)~,
\eeq
and is positive for all type II and type III momentum modes. With these considerations, we find, for $z_i\leq z\leq z_2$,
\beqa \label{eqA5}
\log n_{k}^{res}(z)&={1\over2a}\bigg[(2az'+b)\mu_k(z')+\nonumber\\
&+{d^2\over\sqrt{-a}}\arcsin\bigg({2az'+b\over d}\bigg)\bigg]_{z_i}^{z}~.
\eeqa

Taking into account that $\mu_k(0)=\sqrt{c}/2$, $\mu_k(z_1)=\mu_k(z_2)=0$ and also that $(2az_1+b)/d=-(2az_2+b)/d=1$, we can write the total particle number produced by the resonance for type II modes as:
\beq \label{eqA6}
\log n_{k}^{res\:(II)}(z_2)=-{1\over4a}\bigg[b\sqrt{c}+{d^2\over2\sqrt{-a}}\bigg({\pi\over2}+\arcsin\bigg({b\over d}\bigg)\bigg)\bigg]~.
\eeq
Similarly, for type III modes, we obtain
\beq \label{eqA7}
\log n_{k}^{res\:(III)}(z_2)={\pi\over8}{d^2\over(-a)^{3\over2}}~.
\eeq

It is easy to check that Eqs. (\ref{eqA6}) and (\ref{eqA7}) give the same result for the mode in the transition between types II and III, at $\beta=1$, as expected. Recalling that $|q_0|\ll1$, these expressions can be approximated by those given in Eq. (\ref{eq101}).

\vfill



\end{document}